\documentstyle[twoside,epsf,12pt]{article}
\makeatletter
\@addtoreset{equation}{section}
\makeatother

\newcommand{\bibi}{\bibitem}

\def\a{\alpha}

\def\d{\delta}
\def\e{\epsilon}   
\def\f{\varphi}    

\def\j{\psi}
\def\k{\kappa}     

\def\m{\mu}

\def\n{\nu}

\def\p{\pi}       
\def\t{\tau}

\def\x{\xi}

\def\D{\Delta}

\def\G{\Gamma}

\def\S{\Sigma}

\def\jb{\overline{\j}}

\def\Ds{D\!\!\!\!/\,}

\newcommand{\del}{\partial}

\newcommand{\half}{\mbox{{\normalsize $\frac{1}{2}$}} }

\newcommand{\quart}{\mbox{{\small $\frac{1}{4}$}} }
\newcommand{\twothird}{\mbox{{\small $\frac{2}{3}$}} }

\newcommand{\eigth}{\mbox{{\normalsize $\frac{1}{8}$}} }

\newcommand{\fthird}{\mbox{{\small $\frac{4}{3}$}} }

\newcommand{\ra}{\rightarrow}

\newcommand{\aplt}{ \mbox{}_{\textstyle \sim}^{\textstyle < }     }
\newcommand{\apgt}{ \mbox{}_{\textstyle \sim}^{\textstyle > }     }

\newcommand{\lag}{\langle}
\newcommand{\rag}{\rangle}

\newcommand{\th}{\theta}

\newcommand{\tk}{\widetilde{\kappa}}

\newcommand{\ph}{\phi}

\newcommand{\nnn}{\nonumber \\}

\newcommand{\ds}{\del\!\!\!/}

\newcommand{\hmu}{\hat{\mu}}
\newcommand{\hnu}{\hat{\nu}}

\newcommand{\be}{\begin{equation}}
\newcommand{\ee}{\end{equation}}
\newcommand{\bea}{\begin{eqnarray}}
\newcommand{\eea}{\end{eqnarray}}
\newcommand{\eq}{\ref}
\newcommand{\beq}{\begin{equation}}
\newcommand{\eeq}{\end{equation}}
\newcommand{\cc}{\cite}
\newcommand{\lb}{\label}

\def \3{\ss}

\def\footnoteitem(#1)#2{
\begin{list}{#1}{\labelwidth4.0mm \leftmargin7.0mm
\labelsep2.5mm \rightmargin7.0mm \parsep0.5ex plus0.2ex minus0.1ex
\itemsep0ex plus0.2ex }
\item #2
\end{list}
}

\begin{document}


\headsep=0.0cm
\vsize 25.0truecm
\topmargin=0cm
\topskip=0cm

\begin{titlepage}

\rightline{TAUP-2447-97}
\rightline{Wash. U. HEP/97-61}
\rightline{HU-EP-97/46}
\rightline{UTCCP-P-24}
\vskip 3mm
\rightline{August 1997}

\baselineskip=20pt plus 1pt
\vskip 0.5cm

\centerline{\LARGE On the Phase Diagram of a Lattice U(1) Gauge Theory }
\vskip 0.5cm
\centerline{\LARGE with Gauge Fixing}
\vskip 1.3cm
\centerline{\large Wolfgang Bock$^\#$, Maarten F.L. Golterman$^\$$,
Yigal Shamir$^\& $}
\vskip 2.7cm
\baselineskip=12pt plus 1pt
\parindent 20pt
\centerline{\bf Abstract}
\textwidth=6.0truecm
\medskip

\frenchspacing
As a first step towards a nonperturbative investigation
of the gauge-fixing (Rome) approach to lattice chiral gauge theories we study
a U(1) model with an action that includes a local gauge-fixing term and a mass
counterterm for the gauge fields. The model is studied on the trivial orbit
so that  only the dynamics of the longitudinal
gauge degrees of freedom is taken into account.
The phase diagram of this higher-derivative scalar field
theory is determined, both in the mean-field approximation and numerically.
The continuum limit of the model corresponds to a continuous phase
transition between  a ferromagnetic (FM) phase where the global 
U(1) symmetry is broken, and a 
so-called helicoidal ferromagnetic (FMD) phase
with broken U(1) symmetry and 
a nonvanishing condensate of the vector field. The global U(1) symmetry 
is restored in this continuum limit.
We show that our data for the magnetization in the FM and FMD phases
are in good agreement with perturbation theory.
\nonfrenchspacing

\vskip 1.2cm
\noindent $^\#$ Institute for Computational Physics, Humboldt University Berlin \\
\noindent \phantom{$^\#$} Invalidenstr. 110, 10099 Berlin, Germany \\
\noindent \phantom{$^\#$} e-mail: {\em bock@linde.physik.hu-berlin.de}  \\
\vskip 0.5cm
\noindent $^{\$}$ Department of Physics, Washington University \\
\noindent \phantom{$^\$$} St. Louis, MO 63130, USA\\
\noindent \phantom{$^\$$} e-mail: {\em maarten@aapje.wustl.edu}  \\
\vskip 0.5cm
\noindent $^{\&}$ School of Physics and Astronomy,  \\
\noindent \phantom{$^{\&}$} Beverly and Raymond Sackler Faculty of Exact Sciences
Tel-Aviv University \\
\noindent \phantom{$^{\&}$} Ramat Aviv 69978, Israel \\
\noindent \phantom{$^{\&}$} e-mail: {\em ftshamir@wicc.weizmann.ac.il}  \\

\end{titlepage}

\textheight=20.5cm
\headsep=2.0cm
\vsize 21truecm
%
\section{Introduction}
\lb{INTRO}
All existing lattice fermion formulations   
have in common that they are in conflict with chiral gauge invariance. 
It is well known that for example the Wilson-term \cc{Wi75}, 
which is used to remove the 15 unwanted species doublers at the
corners of the four dimensional Brillouin zone, is not invariant under 
chiral gauge transformations, because it has the structure of
a mass term. 
 
Most fermion formulations can be rendered gauge invariant by inserting 
Higgs fields. The Wilson mass term for instance turns into a Wilson-Yukawa term 
which is invariant under chiral gauge transformations \cc{Sm80,Sw84}. 
These Higgs fields do not need to be 
added by hand. They appear automatically in the gauge noninvariant 
model when performing the 
integration over all gauge fields in the lattice path integral 
with the Haar measure \cc{FoNi80}. The Higgs fields can be identified with 
the longitudinal gauge degrees of freedom. 

The failure of  most proposals for lattice chiral gauge theories, 
such as the Eichten-Preskill model \cc{GoPe93}, the domain wall fermion 
model with waveguide \cc{GoJa94} or the Wilson-Yukawa (Smit-Swift)
model \cc{GoPe92} is connected to the dynamics of these Higgs fields. 

In the Wilson-Yukawa model, for example, it has been shown, that the 
15 unwanted species doublers 
can indeed be removed from the spectrum by means of the Wilson-Yukawa term in 
the strong Wilson-Yukawa coupling symmetric phase. The model  
fails however because the left-handed fermion, 
which is the one that is supposed to couple to the gauge field, 
forms a fermionic bound state 
with the Higgs field that does not 
transform under the gauge group. This neutral 
left-handed fermion pairs up with the right-handed fermion to form 
a Dirac fermion which in the continuum limit decouples from the 
gauge field \cc{GoPe92}.  Later, arguments have been given that the spectrum in a 
symmetric phase is in general vector-like \cc{Sh93,Sh96r}. 

The details of the mechanism which spoils the chiral nature 
of the fermions differ from model to model,
but remarkably the Higgs fields (longitudinal gauge degrees of freedom)
play the key role (for recent reviews 
see refs.~\cc{Sh96r,Pe93}).
It is therefore natural that one should try to use gauge fixing 
to control the effect of these longitudinal gauge degrees of freedom \cc{Sh95}.  

Gauge fixing has been put forward some time ago as a method to 
discretize chiral gauge theories on the lattice \cc{BoMa89}. 
It was proposed in ref.~\cc{BoMa89}  
to use perturbation theory in the continuum as 
guideline and transcribe the gauge-fixed continuum path integral 
to the lattice. Since the fermion part in the action breaks 
chiral gauge invariance, the lattice model is not invariant under BRST 
symmetry. The symmetry breaking terms 
are compensated by counterterms that have to be 
added to the action. In four dimensions  
only the counterterms with dimension smaller than
or equal to four need to be considered. They should furthermore respect 
all exact symmetries of the lattice model.                   
The coefficients of these terms then have to be adjusted such that BRST 
invariance is restored in the continuum limit. 
A lattice discretization of a 
nonlinear gauge, $\sum_\m \{ \del_\m A_\m +  A_\m^2 \} =0$, has been  
introduced in ref.~\cc{Sh95}, while a lattice discretization of 
the Lorentz gauge $\sum_\m \del_\m A_\m=0$ was later given in ref.~\cc{GoSh96}. 
It was pointed out in ref.~\cc{Sh95} that the lattice action 
of the gauge-fixing approach can be rewritten with Higgs fields and 
that on the trivial orbit                         
the gauge-fixing term reduces to a higher-derivative scalar field theory.          
Higher-derivative scalar field theories have been studied recently in 
a series of papers \cc{JaKu93}, but in a very different context.

The Higgs field represents the longitudinal modes of the gauge field.
The central question is then whether the fluctuations of the 
longitudinal modes {\em alias} the Higgs field are sufficiently 
reduced by the gauge-fixing term that the chiral nature of the fermions
does not get spoiled \cc{Sh95}.  In order to study this question, it is
useful to introduce the notion of a {\em reduced model}. The reduced model is
obtained by keeping only
the Higgs fields, setting the gauge field equal to zero.
(In the model without explicit Higgs field, this corresponds to restricting the
gauge field to the trivial orbit.)  This reduced model should then have a 
continuum limit with free fermions in the desired chiral representations
of the gauge group.

In this paper, we will consider the reduced version of the purely bosonic
model with  compact U(1) symmetry that apart from the usual plaquette term
includes the Lorentz gauge-fixing term of ref.~\cc{GoSh96}
and a mass counterterm for the gauge
field.  We will derive the phase diagram of the reduced model, using 
mean-field and numerical techniques.  We will demonstrate the existence of
a continuous phase transition between a phase with broken symmetry (FM phase)
and a 
so-called {\em helicoidal ferromagnetic} (FMD) phase, where a continuum limit
can be defined.  The nature of this continuum limit
will be investigated using both perturbation theory and numerical simulations.
In the framework of the Smit-Swift model we will demonstrate in two
forthcoming publications \cc{BoGoSh97b,BoGoSh97c}
that the unwanted Higgs-fermion bound states
do indeed not emerge, when fermions are coupled  
to this model. The spectrum of this fermion-Higgs
model contains only a left-handed fermion which
couples to the gauge field, if we would turn it on again,
and a right-handed free fermion
which decouples from the gauge field in the continuum limit.

The outline of the paper is as follows: In Sect.~\ref{MODEL} we 
start from the full action of 
the gauge-fixing (Rome) approach and reduce it step by step to the 
``reduced model" which   
will be the subject of this paper. In Sect.~\ref{CMFA} we determine the 
phase diagram of the reduced model in the mean-field approximation.
The magnetization is calculated in the weak coupling expansion 
in Sect.~\ref{PERT}. We demonstrate that the magnetization 
vanishes when approaching the FM-FMD phase transition from 
both the FM and FMD phases. 
Sect.~\ref{NUM} deals with the results of the numerical simulation. The 
numerical results for the phase diagram are presented and 
compared with the mean-field results in Subsect.~\ref{NUM_PD}. The perturbative 
results for the magnetization are compared with numerical data of 
high precision in Subsect.~\ref{CPERT}. 
A brief summary of our results and an outlook 
is given in Sect.~\ref{OUT}.
\section{The Model}
\lb{MODEL}
In the continuum, gauge fixing is needed in order to make the integration over gauge 
orbits well-defined. The continuum  gauge-fixed action for 
a chiral gauge theory can be written as                    
\be
S_{\rm c}=S_{{\rm c, G}}(A_\m^a)+S_{\rm c, F}(A_\m^a; \j_L, \j_R) + S_{{\rm c, g.f.}}(A_\m^a) 
+ S_{{\rm c, ghost }}(A_\m^a; \overline{c}, c) \;.  \lb{CACTION}
\ee
Here  
\be
S_{{\rm c, F}}(A_\m^a; \j_L, \j_R) = 
\int d^4 x \; \left\{ \jb_L(x) \; \Ds \; \j_L (x) 
+ \jb_R(x) \; \ds \; \j_R (x) \right\} \lb{CSF}
\ee
is the fermionic part of the action. Only the left-handed component of 
the fermion couples to the gauge field.
$S_{{\rm c, G}}(A_\m^a)$ is the gauge action,  
$S_{{\rm c, g.f.}}(A_\m^a)$
the gauge-fixing action and $S_{{\rm c, ghost }}(A_\m^a; \overline{c}, c)$ 
is the Fadeev-Popov ghost action. For the Lorentz gauge,               
\bea
S_{{\rm c, g. f.}}(A_\m^a )    \!\!&=&\!\!  
\frac{1}{2 \x} \sum_a \left( \sum_\m \del_\m A_\m^a \right)^2\;, \lb{CSGF}  \\
S_{{\rm c, ghost }}(A_\m^a ; \overline{c} , c)\!\! &=&\!\!  
\sum_{a,b} \overline{c}_a 
\left[ \d_{a,b} \; \Box^2 + g \; f_{abc} \; A^c_\m  \; \del_\m \right] c_b \;,  
\lb{CSGHOST} 
\eea
where $c$ designates the complex ghost field,
$\x$ is the gauge-fixing parameter, $g$ the gauge coupling, $A_\m^a$ the 
gauge field and $f_{abc}$ are the 
structure constants of the gauge group.
The continuum path integral is invariant under BRST transformations which 
replace the local gauge invariance. 

The action is transcribed to the lattice using the compact 
lattice link variables 
\be
U_{\m x} =\exp(i a g  A_{\m x}) \in G  \;, \lb{UUU}
\ee
where $a$ denotes the lattice spacing. In the following we will  
set $a$ equal to one.       
The compact link variables are elements of the gauge group  
G and are assigned to the lattice links $(x, x+\hmu)$. The action on the 
lattice can then be written in the form, 
\be
S=S_{\rm G}(U)+S_{\rm F}(U; \j_L, \j_R) + S_{{\rm g. f.}}(U) 
+ S_{{\rm  ghost }}(U ;\overline{c}, c) 
+S_{{\rm c.t.}}(U;\j_L, \j_R;\overline{c} , c)\;.
\lb{FULL_ACTION}
\ee
Only the plaquette action $S_{\rm G}(U)$ is manifestly 
gauge invariant. To transcribe the fermion action (\eq{CSF}) to the lattice one 
has to choose a particular lattice fermion formulation (like
Wilson, domain wall or staggered fermions). 
All known lattice fermion formulations are in conflict with 
local chiral gauge invariance and, as a consequence,
$S_{\rm F}(U; \j_L, \j_R)$ is not  invariant under BRST symmetry. 
This means that counterterms have to  be added to the action 
so that this symmetry is restored in the continuum limit. We have to consider all 
terms with  dimension smaller than or equal to four, which respect 
all exact symmetries of the lattice model. 

In this paper we will not consider fermions and focus only on the 
discretization of the bosonic part of the action  (\eq{CACTION}).
It turns out that also 
the gauge-fixing part of the action,  i.e. the combination
$ S_{{\rm g. f.}}(U)+ S_{{\rm  ghost }}(U ;\overline{c}, c) $, 
should be formulated on the lattice such that BRST symmetry 
is broken. The reason is a theorem \cc{Ne87} 
which states that the partition function itself, as well as
expectation values of gauge invariant observables, vanish
in a lattice model with exact BRST invariance, due to the existence of lattice
Gribov copies (see also ref. \cc{Sh84}).

As a second simplification we choose U(1) as gauge group. This 
choice makes both the analytical and numerical 
calculations considerably easier since the ghost action in eq.~(\eq{CSGHOST})
does not depend on the gauge potential and there         
are also no counterterms which couple the ghosts to other fields in the action.
This implies that in the abelian case
the ghost sector can be dropped  completely from the path integral.

Finally, as a third simplification we 
include only the gauge-boson mass counterterm for the 
gauge field (this is the only counterterm of 
dimension two) and ignore all counterterms of higher dimension.     
The coefficient of this mass counterterm has to be tuned such that the photons are 
massless.  For all dimension-four counterterms without derivatives 
it has been argued         
in ref.~\cc{GoSh96} that they do not  alter                   
the phase structure of the physically relevant region 
of the phase diagram (the existence of a continuous FM-FMD phase transition).
We believe that this remains also true if all other 
counterterms are included in the action.

 The U(1) model we are studying in this paper           
is then defined by the path integral, 
\be
Z=\int D U \; \exp( -S(U))   \; \lb{PATH0}
\ee    
where the action is given by 
\bea
S(U) \!\!&=&\!\! S_{{\rm G}}(U) + S_{{\rm g. f.}}(U) + S_{{\rm m}} (U)\;, \lb{S} \\
S_{{\rm G}}(U) \!\!&=&\!\! \frac{1}{g^2} \; \sum_{x \m \n} 
\left\{ 1-\mbox{Re } U_{\m \n x} \right\} \;,  \lb{SG} \\
S_{{\rm g. f.}}(U) \!\!&=&\!\! \tk \left\{
\sum_{x,y,z} \Box (U)_{xy} \; \Box (U)_{yz} - \sum_x B_x^2 \right\} \; , 
  \lb{SGF} \\
S_{{\rm m}} (U)\!\!&=&\!\! -\k  \sum_{\m x} \left\{ U_{\m x}+ U_{\m x}^{\dagger}
\right\} \;. \lb{SM}
\eea
Here $U_{\m \n x}=U_{\m x} U_{\n x+\hmu} 
U_{\m x+\hnu}^{\dagger} U_{\n x}^{\dagger}$ 
is the usual plaquette variable, 
\be
\Box (U)_{xy} = \sum_\m \left\{ U_{\m x}\; \d_{x+\hmu,y} 
			       + U_{\m x-\hmu}^{\dagger}\; \d_{x-\hmu,y} 
			       -2 \; \d_{x,y} \right\} \lb{LL}
\ee
is the covariant lattice laplacian, 
\be
B_x=\sum_\m \left( \frac{V_{\m x-\hmu} + V_{\m x} }{2} \right)^2 \;, \lb{BX}
\ee
with 
\be
V_{\m x} = \frac{1}{2i} \; \left( U_{\m x} - U_{\m x}^{\dagger} \right) 
= g \; A_{\m x} + O( (g \; A_{\m x})^3) \;, 
\lb{V}
\ee
and 
\be
\tk= \frac{1}{2 \x g^2 }\;.  \lb{TKG2}
\ee
The reader can easily verify that the gauge-fixing 
action (\eq{SGF}) reduces in the classical
continuum limit to eq.~(\eq{CSGF}). The gauge-fixing term  (\eq{CSGF})
can be transcribed to the lattice in many different ways.
The choice in eq.~(\eq{SGF}) is motivated by the following 
important properties \cc{GoSh96}:
\begin{enumerate}
\item The action (\eq{SGF}) has a unique absolute minimum at $U_{\m x}=1$, 
      validating the weak coupling expansion.  
      
\item  The action (\eq{SGF}) 
      is not BRST invariant. This is related to the fact that it cannot be 
      written as a square of (a discretized version of)
      the gauge-fixing functional, $\sum_\m \del_\m A_\m $.              
      The theorem of ref.~\cc{Ne87} therefore does not apply in our case.

\item The action (\eq{SGF}) leads to critical behavior in 
      the continuum limit $g \ra 0$. 
\end{enumerate}
In connection with item 1 we 
note that the naive discretization of the gauge-fixing action 
\be
S_{{\rm g.f.}}(U) = \frac{1}{2 \x g^2 } 
\sum_x  \left( \sum_\m \left( V_{\m x} - V_{\m x-\hmu} \right) \right)^2 \;,
\ee
where eq.~(\eq{V}) was used to transcribe $A_{\m x}$ in (\eq{CSGF})
in terms of $U_{\m x}$, does not have a unique minimum. 
It was demonstrated in 
ref.~\cc{Sh95} that this action gives rise to a dense set of lattice 
Gribov copies. Such a dense set of Gribov copies may still give rise to strong 
fluctuations of the longitudinal gauge degrees of freedom, 
a situation which we want to avoid from the start.               

\begin{figure}
\centerline{
\epsfysize=14.0cm
\vspace*{0.5cm}
\epsfbox{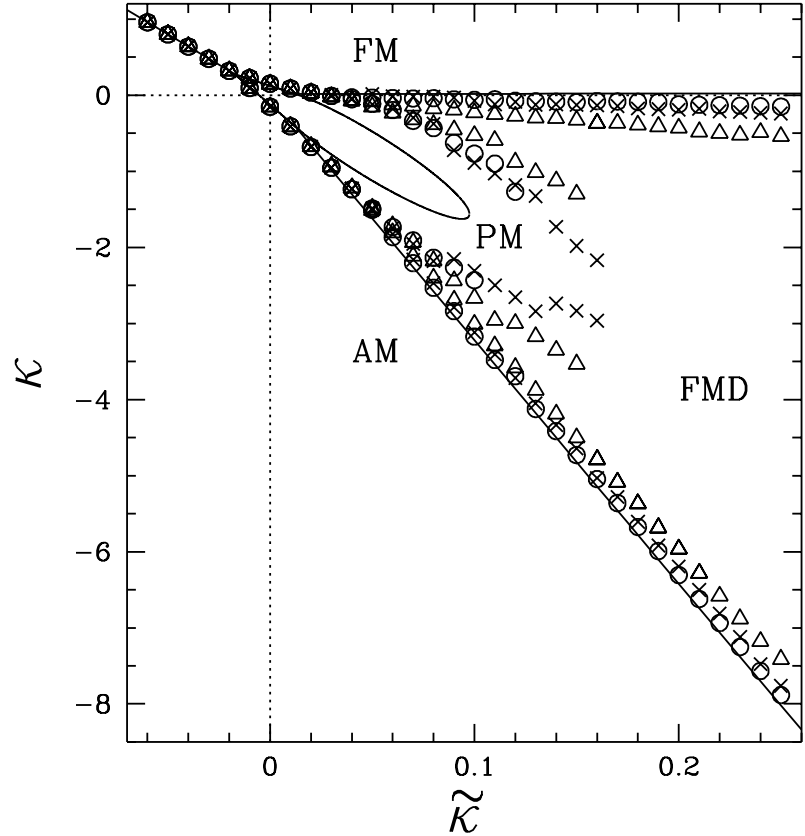}
}
\vspace*{-1.2cm}
\caption{ \noindent {\em  The $(\k, \tk)$ phase diagram of the reduced model 
({\protect \eq{ACTION}}) 
contains four different phases: a ferromagnetic (FM), an antiferromagnetic 
(AM), a paramagnetic (PM) and a directional ferromagnetic (FMD) phase.
The phase boundaries in the mean-field approximation,  
obtained in the infinite volume limit, are represented  
by the solid lines. 
The numerical results for the phase transitions, 
obtained by scanning the parameter space in $\k$ direction on the 
$4^4$, $6^4$ and $8^4$ lattices are marked by the triangles, crosses and 
circles. The error bars are omitted in all cases since they 
are smaller than the symbol size.
}}
\label{phase_diagram}
\end{figure}
 Some first information about the phase structure of the model  
is obtained in the constant field approximation. In this approximation 
the  lattice link 
field in eq.~(\eq{UUU}) is replaced by a constant gauge field that is independent 
of $x$. All terms which contain derivatives of the gauge field vanish when 
we insert this constant gauge field into eq.~(\eq{S}) and therefore 
we obtain an expression for the classical potential.
After expanding the resulting expression for the classical potential 
in powers of $g$ we find 
\be
V_{{\rm cl}}(A_\m)= \k \left\{ g^2 \sum_\m A_\m^2 + \ldots \right\} + \frac{g^4 }{2 \x} \left\{ \left( \sum_\m A_\m^2 \right)
\left( \sum_\m A_\m^4 \right) +\ldots \right\} \;, \lb{VCL}
\ee
where the $\dots$ represent terms of higher order in $g^2$. The coefficient of 
the term quadratic in $A_\m$ has to be tuned such 
that the gauge-boson mass vanishes. The value of $\k$ where     
the gauge-boson mass vanishes at a given value 
of $g^2$ defines a critical point $\k_{{\rm FM-FMD}}(g^2)$. Using         
eq.~(\eq{TKG2}) we can also replace $g^2$ everywhere in the series by $\tk$.
At tree-level we find, from eq.~(\eq{VCL}),
\be
\k_{{\rm FM-FMD}}(\tk)=0\;, \;\;\; \lb{TREE}
\ee
which is a good approximation only in the large $\tk$ region of the 
phase diagram. A first glance at the phase diagram in 
fig.~\ref{phase_diagram} (which 
was obtained on the trivial orbit, i.e. $U_{\m x}=g_x g_{x+\hmu}^{\dagger}$)
shows that the mean field and numerical results for 
the FM-FMD phase transition
at large $\tk$ are indeed very close to the $\k=0$ axis.
We will show later that the critical coupling 
$\k_{{\rm FM-FMD}}(\tk)$ is shifted by perturbative 
corrections to a small positive value.                        

The minimization of eq.~(\eq{VCL}) shows that 
\be
\begin{array}{ll}
\lag  g \; A_\m \rag = 0\;, &
 \mbox{for } \k \geq \k_{{\rm FM-FMD}} \;,  \\            
\lag g \; A_\m \rag = \pm \left( |\k -\k_{{\rm FM-FMD}}| / (6 \; \tk)  \right)^{1/4} 
\;,  &  \mbox{for } \k < \k_{{\rm FM-FMD}} \;, 
\end{array}
\lb{CFMD} 
\ee
for all $\m=1,\ldots,4$ \cc{GoSh96}. 
This implies that $\k=\k_{{\rm FM-FMD}}=0$ corresponds to a phase transition between 
a ferromagnetic (FM) phase, where $\lag A_\m \rag$ vanishes and  the gauge boson
has a nonzero mass, and  
a so-called FMD phase with a nonvanishing 
vector condensate $\lag A_\m \rag$. It will become clear shortly why these phases 
are called 
ferromagnetic. The abbreviation FMD stands for {\em ferromagnetic directional}
to express the fact that the vectorial vacuum 
expectation value $\lag A_\m \rag$ induces a  space-time direction 
and that, as a consequence, hypercubic rotation invariance is broken. 
The continuum limit of the model corresponds to the continuous phase
transition between  the FM phase and the FMD phase.

To investigate the properties of the model (\eq{S}) beyond the constant 
field approximation we can study fluctuations around the classical ground state 
in the FM and FMD phase by expanding the observables in powers of $g^2$, 
or alternatively powers of $1/\tk$, cf. eq.~(\eq{TKG2}). 

Next we demonstrate that the model            
defined by the path integral (\eq{PATH0}) is equivalent to 
a gauge-Higgs model which is manifestly gauge invariant.
The lattice path integral in (\eq{PATH0}) can be rewritten as      
\cc{FoNi80}
\bea
Z\!\!&=&\!\!\int D U \; \exp( -S(U_{\m x}))  \nnn
\!\!&=&\!\! \int D U 
\; \exp( -S( g_{x} U_{\m x} g_{x+\hmu}^{\dagger}))  \nnn 
\!\!&=&\!\! \int D U D \phi \;
   \exp( -S(\phi_x^{\dagger} U_{\m x} \phi_{x+\hmu} ))  \lb{PATHI}
\eea
where in the second line we have performed a local gauge transformation,  
$U_{\m x} \ra  g_x U_{\m x} g_{x+\hmu}^{\dagger}$, 
$g_x \in$G and made use of the gauge invariance of the Haar measure. 
The $g_x$'s drop out of the plaquette action 
because it is manifestly gauge invariant. The 
third line is obtained after integrating both sides of the 
equation (\eq{PATHI}) 
over the gauge degrees of freedom with $g_x=\phi_x^{\dagger}$ and using 
the fact that $\int D \phi=1$. This simple transformation 
shows that the longitudinal gauge degrees of freedom turn into group-valued Higgs 
fields. The new action 
$S(\phi_x^{\dagger} U_{\m x} \phi_{x+\hmu}) $  is now 
invariant under the gauge transformations 
\be
U_{\m x} \ra h_x U_{\m x} h_{x+\hmu}^{\dagger} \;, \;\;\;\; 
\phi_x \ra  h_x \phi_x  \;. \lb{GT}
\ee
We will refer in the following to $S_V=S(U_{\m x})$ in eq.~(\eq{S})
as the action in the {\em vector} picture and to
$S_H=S(\phi_x^{\dagger} U_{\m x} \phi_{x+\hmu})$ in eq.~(\eq{PATHI})
as the action in the {\em Higgs} picture.  The two actions 
are related by 
\be
S_V(U_{\m x}) =\left. S_H(U_{\m x}; \; \phi_x) \right|_{\phi=1}
\ee
and all observables in the vector picture are mapped onto 
corresponding observables in the Higgs picture (see also ref. \cc{Sh95}).

In this paper we will study a {\em reduced} model defined 
by the action (\eq{S}) on the trivial orbit,                     
$U_{\m x}=g_x \; 1\; g_{x+\hmu}^{\dagger}$. 
In the Higgs picture, cf. eq.~(\eq{PATHI}), the reduced model is obtained by 
setting $U_{\m x}=1$. The reduced model is then defined by 
the following lattice path integral 
\bea
Z\!\!&=&\!\!\int D \phi  \; \exp( -S(\phi) ) \lb{PATH} \\    
S(\phi )\!\!&=&\!\! -\k \; \sum_x \phi^{\dagger}_x (\Box \phi)_x 
                    +\tk \; \sum_x \left\{ \phi^{\dagger}_x (\Box^2 \phi)_x -
		    B_x^2 \right\} \;, \lb{ACTION}
\eea
where $B_x$ is given by eq.~(\eq{BX}) with 
\be
V_{\m x} =\frac{1}{2i} \; \left( \phi_x^{\dagger} \; \phi_{x+\hmu}-
\phi_{x+\hmu}^{\dagger} \; \phi_{x} \right) \;. \lb{VMX}
\ee
Eq.~(\eq{ACTION}) defines a higher-derivative scalar field theory. 
       
As a first step we will investigate in the following 
the phase diagram of the reduced model. Eq.~(\eq{ACTION}) shows that 
the partition function is invariant under the symmetry 
\be
\k \ra -\k -32\; \tk \;, \;\;\;\;\; 
\tk \ra \tk   \;, \;\;\;\;\;
\phi_x \ra \epsilon_x \; \phi_x \;, \lb{SYMM}   
\ee
where
\be
\epsilon_x=(-1)^{\Sigma (x)} \;,\;\;\;\;\;\;\;\;\;\Sigma (x)=\sum_\m x_\m \;. \lb{SIGM}
\ee
This implies that the phase diagram is symmetric under reflection 
with respect to the line 
\be
\k +16 \; \tk=0 \;.  \lb{LINE}
\ee

For $\tk=0$ we recover the XY model in four dimensions 
whose phase diagram consists of three different phases: a 
broken or ferromagnetic (FM) phase at $\k > \k_{{\rm FM-PM}}>0$, 
a symmetric or paramagnetic (PM) 
phase at $\k_{{\rm PM-AM}} < \k < \k_{{\rm FM-PM}}$  
and an antiferromagnetic (AM) phase 
at $\k< \k_{{\rm PM-AM}}<0$. The symmetry (\eq{SYMM}) implies 
that,  $\k_{{\rm PM-AM}}=-\k_{{\rm FM-PM}}$. Numerically it has been 
found that $\k_{{\rm FM-PM}} \approx 0.15$.
The order parameters which allow us to distinguish between 
these phases are the magnetization 
\be
v= \left| \left \langle \phi_x \right\rangle   \right|  \lb{YMAG}   
\ee
and the staggered magnetization 
\be
v_{{\rm AM}}= \left| \left \langle  \epsilon_x \; \phi_x \right\rangle 
\right| \;.
\lb{YMAGST}            
\ee
Both quantities are not invariant under the global U(1) symmetry,       
and we have taken the modulus to eliminate the ambiguity 
due to the constant field mode.
The FM phase is characterized by $v > 0$, $v_{{\rm AM}} = 0$, 
whereas in the AM phase $v= 0$, $v_{{\rm AM}} > 0$. 
Both order parameters vanish in the intermediate PM phase.  

As explained above, at large $\tk$ we expect to find a new phase transition 
between the FM and the FMD phase, which 
at tree level is given by eq.~(\eq{TREE}).
(In the following we will  retain    
the name FMD also for the reduced-model version of the FMD phase.) The 
FMD phase is characterized by a new vector order parameter $q_\m$, $0< q_\m < 2\pi$,
which is nonzero in the FMD phase and vanishes in the FM phase. It is equal  
to $(\pi,\pi,\pi,\pi)$ in the AM phase. As a generalization of $v$ and 
$v_{{\rm AM}}$ we define a helicoidal magnetization 
\be
v_{{\rm H}} 
= \left| \left\lag  \phi_x \; \exp \left( -i \; \sum_\m q_\m \; x_\m \right)
\right\rag  \right| 
\;, \lb{YDMAG}            
\ee
which is nonzero in the FMD phase. It is easy to see that $v_{{\rm H}}$
reduces to $v$ in the FM and to $v_{{\rm AM}}$ in AM phase.

When ignoring fluctuations around the ground   
state, the vector field $V_{\m x}$, cf. eqs.~(\eq{V}) and (\eq{VMX}), 
in the FMD phase is given by 
\be
V_{\m x} = v_{{\rm H }}^2 \; q_\m + O( q_\m^3 ) \;,  \lb{VV}            
\ee
showing that $q_\m$ plays the role of the vector condensate in the reduced model.
We mention in passing that phases with nonvanishing $q_\m$ have been intensively 
investigated in lower dimensions in condensed matter 
physics and are known as helicoidal-ferromagnetic phases (see ref.~\cc{Se92} 
for a recent review).

To further substantiate the statements about the phase diagram 
made in this section we 
will determine in the next section the phase diagram
of the reduced model (\eq{ACTION}) in the mean-field approximation. Numerical 
data for the phase diagram 
are presented in Sect.~\ref{NUM_PD} and  compared with the 
mean-field results. 
\section{The Phase Diagram in the Mean-Field Approximation}
\lb{CMFA}
In the following we will 
perform a mean-field analysis of the phase diagram in $d$ dimensions. 

A central problem of the mean-field approximation in more complicated 
ferromagnetic systems is the choice of the mean-field 
ansatz which in a given 
region of the parameter space leads to the absolute minimum of the free energy.
Usually there exist many different choices 
and it is not straightforward to pick an ansatz which leads 
to the absolute minimum of the free energy. 
Based on the discussion of the previous section we decided to  
consider the ansatz 
\be
\ph_x= \f \; \exp \left(i \; \sum_\m  q_\m  \; x_\m \right)\;, \lb{MFA}
\ee
where $q_\m$,  $0 \leq q_\m < 2\p$ are real phases and 
$\f$ plays the role of a magnetization. Depending on the 
value of $q_\m$ this ansatz can distinguish between phases with 
ferromagnetic  ($q_\m=0$, $\m=1,\ldots,d$), $\f=v$, 
antiferromagnetic ($q_\m=\p$, $\m=1,\ldots,d$) 
ordering, $\f=v_{{\rm AM}}$, and phases with a helicoidal magnetization
($q_\m \neq 0,\p$, for at least one component $\m$), $\f=v_{{\rm H}}$. 
Similarly, we take for the magnetic field $h_x$ the ansatz
\be
h_x= h \; \exp \left(i \; \sum_\m  q_\m  \; x_\m \right)\;, \lb{MFAH}
\ee
where $h$ is the mean-field magnetic field.

Using eqs.~(\eq{MFA}) and (\eq{MFAH}) and 
following the steps of the standard mean-field calculation  
(see for example ref.~\cc{Zi96}), 
we obtain for the free energy of the reduced model 
(\eq{ACTION}) 
\be
{\cal F}(\f,h,q;\tk,\k )= 
L^d \; \left\{ 2 \; \f \; h 
- \log \; I_0 (2h)+ \sum_{i=1}^4 \; \f^{2i} \; f^{(i)} (q;\tk,\k) \right\} \;, \lb{FREE}
\ee
where $L$ is the extent of the lattice 
in spatial and temporal directions, 
\bea
 \!\!\!\!\!\! f^{(1)}(q;\tk,\k)\!\!\!&=&\!\!\! -2 \; (4\; d\; \tk +\k) \; F(q)
                          +2 \; \tk \; (2 F(q)^2 -d) + \frac{\tk}{16} 
			  \; F(2q) \; (2d+1)  \;,
			  \lb{F_1} \\
 \!\!\!\!\!\! f^{(2)}(q;\tk,\k)\!\!\!&=&\!\!\! - \frac{\tk}{64} \; \left( 
			   6 \; F(2q)^2 -4 \; (3 \; d+1) \; F(2q) 
			  +d\; (10\; d-1)  \right) \;,
			  \lb{F_2} \\
 \!\!\!\!\!\! f^{(3)}(q;\tk,\k)\!\!\!&=&\!\!\! - \frac{\tk}{16} \; \left( 
			   2 \; F(2q)^2 -2 \; (d-1) \; F(2q) 
			  -d  \right) \;,
			  \lb{F_3} \\
 \!\!\!\!\!\! f^{(4)}(q;\tk,\k)\!\!\!&=&\!\!\! - \frac{\tk}{32} \;
			   ( F(2q) -d )^2   \;,
			  \lb{F_4} 
\eea
\be
F(q)=\sum_\m \cos q_\m  \lb{FQ} 
\ee
and 
\be
I_0(h)=\frac{1}{\p}\int_0^\p d \a \; \exp( \pm h \; \cos \a )
\ee
is the modified Bessel function of zeroth order. We have dropped 
in eq.~(\eq{FREE}) all terms that depend neither on $\f$ nor on 
$q$. The saddle-point equations read 
\bea
\frac{\del {\cal F} }{ \del \f}  \!\!\!&=&\!\!\!
L^d \left\{ 2 \; h + \sum_{i=1}^4 \f^{2i-1} \; 2\; i \; f^{(i)} (q;\tk,\k) \right\} 
= 0  \;, \lb{SPE1} \\
\frac{\del {\cal F} }{ \del q_\m }  \!\!\!&=&\!\!\!
L^d \; \f^2 \; \sin q_\m \; \left\{ \sum_{i=1}^4 \f^{2i-2} \; g^{(i)}_{\m} (q;\tk,\k) \right\} = 0  \;, \lb{SPE2} \\
\frac{\del {\cal F} }{ \del h}     \!\!\!&=&\!\!\! 2\;  L^d \; 
\left\{ \f -\frac{I_1(2h) }{I_0(2h)} \right\}=0  \;, \lb{SPE3}
\eea
where 
\bea
 g^{(1)}_{\m} (q;\tk,\k)\!\!\!&=&\!\!\! 2 \; (4\; d\; \tk +\k) 
                          -8 \; \tk \; F(q) - \frac{\tk}{4} \; (2\; d+1) \; \cos q_\m 
		 	  \lb{FQ_1} \\
 g^{(2)}_{\m} (q;\tk,\k)\!\!\!&=&\!\!\!  \frac{\tk}{16} \; ( 
			   12 \; F(2q) -12 \; d -4) \; \cos q_\m
			  \lb{FQ_2} \\
 g^{(3)}_{\m} (q;\tk,\k)\!\!\!&=&\!\!\! \tk \; ( 
			   F(2q) - \half \; (d-1) ) \; \cos q_\m 
			  \lb{FQ_3} \\
 g^{(4)}_{\m} (q;\tk,\k)\!\!\!&=&\!\!\!  \frac{\tk}{4} \;
			   ( F(2q) -d)  \; \cos q_\m  \;,
			  \lb{FQ_4}  
\eea
and 
\bea
I_1(h)=\frac{d I_0(h)}{d h}
\eea
is the modified Bessel function of first order.

{}From these $d+2$ equations we can compute the $d+2$ fields  
$\f$, $q_\m$ and $h$ as functions of the parameters $\tk$ 
and $\k$. The phase boundaries are defined as the lines 
in the ($\tk$, $\k$) 
parameter space where various combinations of the 
order parameters $\f$ and $q_\m$ vanish. 

The variable $h$ can be eliminated from the saddle-point 
equations in regions where $\f$ is very small (which is the case 
close to the PM phase, where $\f$ vanishes) and 
the ratio $I_1(2h)/I_0 (2h)$ in eq.~(\eq{SPE3}) 
can be expanded in powers of $h$, $I_1(2h)/I_0(2h)=h+O(h^3)$. 

Usually there does not exist a unique solution of 
the saddle-point equations in a certain region of the parameter space. 
It is therefore important to substitute the various solutions back 
into the expression for the free energy (\eq{FREE}) 
and to pick out the solution that corresponds to the absolute minimum.       
In practice it can happen that certain phases remain undetected because 
the mean-field ansatz was too simple. 
In the following we will consider also another ansatz to search for 
a ferrimagnetic (FI) phase 
in a certain region of the parameter space which cannot be probed with the 
ansatz (\eq{MFA}). Because of this uncertainty of the mean-field calculation  
it is important to determine the phase diagram also numerically. 

We furthermore note that the free energy in eq.~(\eq{FREE}) is invariant under 
the symmetry (\eq{SYMM}), 
\be
q_\m \ra \p-q_\m\;,\;\;\;\;\; \k \ra -\k -8\; d \; \tk \;, \;\;\;\;\; 
\tk \ra \tk\;, \lb{F_TRAFO}
\ee
which implies that also the phase diagram in the mean-field approximation 
is symmetric (but for the interchange 
$q_\m \leftrightarrow \p-q_\m$, which maps the FM onto the AM phase, etc.) 
with respect to the line $\k + 4 d \tk=0$ which 
in four dimensions turns into eq.~(\eq{LINE}).                        

In the following paragraph we present our mean-field 
results for the phase boundaries and briefly explain how they   
were obtained: 
\begin{itemize} 
\item {\bf FM-PM and PM-AM transitions:} 
      The transition between the FM and PM (PM and AM) phases 
      is obtained by approaching the transitions from within the FM (AM) 
      phase  where $F(q)=d$, $F(2q)=d$ ($F(q)=-d$, $F(2q)=d$) 
      and $\f=v$ ($\f=v_{{\rm AM}}$) approaches zero.
      The relation for the FM-PM (PM-AM) phase boundary 
      is obtained by expanding eq.~(\eq{SPE1}) in powers of $\f$ and 
      equating the part that is linear in $\f$ with zero. Eq.~(\eq{SPE2}) is 
      trivially fulfilled because $\sin q_\m =0$ in the FM (AM) phases.
      The FM-PM and PM-AM phase boundaries are respectively given by the  
      relations,
\bea
\k_{{\rm FM-PM}} \!\!&=&\!\! 
\frac{1}{2d} 
-(2\;d +1 ) \; 
\frac{31}{32}
 \; \tk  \;,
\lb{FM-PM} \\
\k_{{\rm PM-AM}} \!\!&=&\!\! -
\frac{1}{2d}
-\left( 
\frac{194}{32} \;d - 
\frac{31}{32} 
\right)  \; \tk \;. 
\lb{PM-AM} 
\eea
      The reader can easily verify that the 
      two solutions are related to each other by the transformation (\eq{F_TRAFO}).
      The FM-PM and PM-AM phase transition lines intersect at 
      $\tk =\tk_1= \left[d \; (\frac{31}{16}-\frac{33}{8} \; d \; ) \right]^{-1}$. 
      The corresponding solution of the saddle-point equations minimizes the 
      free energy (\eq{FREE}) only in the interval 
\be
\tk_1 \leq \tk \leq 
\frac{16}{35} \; \left[d \; (2\;d+1) \right]^{-1}=\tk_2\;,
\lb{INT1}
\ee
      and the two straight lines (\eq{FM-PM}) and (\eq{PM-AM}) 
      therefore form the boundary of the PM phase only in that interval.

\item {\bf FMD-PM transition:} The magnetization $\f=v_{{\rm H}}$ vanishes 
      when we approach the phase boundary from the FMD side which means that 
      eqs.~(\eq{SPE1}) and  (\eq{SPE2}) can be expanded in powers of $\f$. 
      Unlike in the 
      previous case $\sin q_\m \neq 0$ for at least 
      one component $\m=1,\ldots,d$ and hence $-d < F(q) < +d$. 
      {}From the term in eq.~(\eq{SPE1}) that is proportional to $\f$ we obtain
\be
2\; (4\;d\;\tk +\k) \; F(q)-2\; \tk \; (2\;F(q)^2-d) 
-\frac{\tk}{16}\; F(2q)\; (2\; d+1)=1 \;. \lb{SPE1_AA}
\ee
       After summing the $i=1$ term 
       inside the curly brackets in eq.~(\eq{SPE2}) over 
      $\m$ we obtain for $F(q)$ the formula
\be
 F(q) = 8\;d\; 
\frac{4\; d\; \tk+\k}{(34\; d + 1) \; \tk }\;,  \lb{SPE2_B}
\ee
      and after first multiplying the same term 
      with $\cos q_\m$ and then 
      summing over $\m$ we obtain an expression for $F(2q)$,
\be
\frac{\tk}{16}\; (2\; d+1) \; F(2q) = (4\;d\;\tk+\k) \; F(q) 
- 4\; \tk \; F(q)^2 - \frac{\tk}{16}\; (2\; d+1) \;d  \;.  \lb{SPE2_A} 
\ee
      After substituting these two solutions into eq.~(\eq{SPE1_AA}) and after a few 
      trivial algebraic manipulations we obtain  the following solution for the 
      FMD-PM phase transition:
\be
\k_{{\rm FMD-PM}} =-4 \;d\; \tk \pm \sqrt{        
\frac{\tk}{8\;d}  \left[1- \frac{\tk}{16} \; d\; (2\;d +33) \right] \; 
(34\;d+1) } \;.
\lb{FMD_PM}
\ee
      This solution describes an ellipse located  
      around the symmetry axis, $\k+4\;d\;\tk =0$. Eq.~(\eq{FMD_PM}) 
      forms the boundary of the PM phase in the interval
\be
\tk_2 \leq \tk \leq 16 \; \left[d\; (2\;d+33) \right]^{-1} =\tk_3 \;,
\lb{INT2}
\ee
      as the corresponding mean-field solution does not lead to an absolute 
      minimum of the free energy (\eq{FREE}) in the region where 
      $\tk \leq \tk_2$.

\item {\bf FM-AM phase transition:} Above we pointed out that 
      the FM-PM and PM-AM phase transitions intersect at 
      $\tk= \tk_1$. This means that 
      the FM and AM meet at this value of $\tk$.
      Two different scenarios are imaginable for the phase structure in the region 
      $\tk < \tk_1$: 
      1) The FM and AM phases meet at the 
      $4\;d\;\tk+\k=0$ symmetry line 
      with the magnetizations $v$ and $v_{{\rm AM}}$
      exhibiting a jump at this line, or 2) the FM and AM phases are 
      separated by a FI phase in which both order parameters 
      $v$ and $v_{{\rm AM}}$ are simultaneously nonzero. 
      The mean-field ansatz (\eq{MFA}) 
      is not suited to detect such an intermediate FI phase since 
      $q_\m$ cannot be equal to 
      $0$ and $\p$ at the same time. We therefore
      calculated the free energy also for the ansatz of the form
\be
\ph_x= v + v_{{\rm AM}} \; \e_x\;,
\lb{MFA_B}
\ee
      which allows us to probe for a FI phase. 
      Our calculation however shows 
      that scenario 2) leads to a larger value of the free energy.
      Also our numerical data  in four dimensions 
      give clear evidence for the correctness of the first scenario.

\item {\bf FM-FMD and FMD-AM transitions:} The FM-FMD (FMD-AM) 
      phase transition is characterized by $q_\m \ra 0$ 
      ($q_\m \ra \p$) for all $\m=1,\ldots,d$. 
      It is difficult to determine the location of 
      these phase transitions 
      analytically because $\f$ does not vanish at these        
      two phase transitions and eqs.~(\eq{SPE1})-(\eq{SPE3}) cannot be 
      expanded in $\f$. We will show in the next section that the 
      magnetization actually vanishes at the FM-FMD (FMD-AM) phase transition.  
      This phenomenon is connected 
      to the infra-red behavior of the higher-derivative action 
      and therefore cannot 
      be understood in the framework of the mean-field approximation. 
      
       The FM-FMD phase transition is determined by eliminating the fields 
      $h$ and $\f$ from equation (\eq{SPE3}) and
\bea
&&\!\!\!\!\!\!\!\!\!\!\!\!\!\! 
  \!\!\!\!\!\!\!
-4\;d \;\k-4\;d\; \tk \; (2\;d+1) +\frac{\tk}{8} \; d \; (2\;d +1) + \frac{1}{16}
\; \tk \; d\; (-4\;d+5)  \; \f^2  \nonumber \\
&&\!\!\!\!\!\!\!\!\!\!\!\!\!\! 
  \!\!\!\!\!\!\!
- \frac{3}{8} \; \tk \; d \; \f^4 +2 \; \frac{h}{\f}=0  \;, \lb{SPE1_C} \\
&&\!\!\!\!\!\!\!\!\!\!\!\!\!\! 
  \!\!\!\!\!\!\!
2\; \k -\frac{\tk}{4} \; (2\; d +1) - \frac{1}{4} \; \tk \; \f^2 
+  \frac{\tk}{2} \; (d+1) \; \f^4 =0 \; . \lb{SPE2_C} 
\eea
      Eqs.~(\eq{SPE1_C}) and (\eq{SPE2_C}) are obtained from eqs.~(\eq{SPE1}) 
      and (\eq{SPE2}) in the limit  
      $q \ra 0$. Note that eqs.~(\eq{SPE3}), (\eq{SPE1_C}) and 
      (\eq{SPE2_C}) can only hold simultaneously on the FM-FMD transition 
      curve, where $F(q)=F(2q)=d$.  The location of         
      the FM-FMD transition can be calculated analytically in two special cases:
      at the PM-phase boundary, the terms in eq. (\eq{SPE2_C}) which 
      are quadratic and quartic in $\f$ can be ignored, and the FM-FMD transition 
      is given by the intersection of the PM-phase boundary and the line 
\be
\k_{{\rm FM-FMD}} = \frac{\tk}{8} \; (2\;d+1) \;. \lb{FM-FMD}
\ee
      Similarly the FMD-AM phase transition is given by 
      the intersection of the PM-phase boundary and the line
\be
\k_{{\rm FMD-AM}} = -8\;d\;\tk -\frac{\tk}{8} \; (2\;d+1)  \;. \lb{AM-FMD}
\ee
       The FM-FMD phase transition can also be calculated analytically in the limit 
      $\tk \ra \infty$. Eq.~(\eq{SPE1_C}) implies that        
      $h= \tk \; d \; (128 \;d +63)/ 32 +O(1)$ for $\tk \ra \infty$, 
      and the expansion of $\f$, cf. eq.~(\eq{SPE3}), gives  
      $\f=1-1/(4h)+O(1/h^2)$. After substituting these two formulas into 
      eq.~(\eq{SPE2_C}) we find 
\be
\k_{{\rm FM-FMD}} \ra \frac{2 \; (4\;d+3)}{(128 \; d + 63)\;d} \; , 
\;\;\;\; \tk \ra \infty   \;. \lb{FM-FMD_IN}
\ee
       In four dimensions we have determined $\k_{{\rm FM-FMD}} (\tk)$ 
      also at a series of intermediate $\tk$ values by solving 
      eqs.~(\eq{SPE3}), (\eq{SPE1_C}) and (\eq{SPE2_C}) numerically.
      The results for $\k_{{\rm FM-FMD}} (\tk)$
      are listed (with an accuracy of four decimal places)
      in the second column of table~\ref{TAB1}. We have also 
      included the mean-field value of $\f$ evaluated at 
      $\k=\k_{{\rm FM-FMD}} (\tk)$ (third column).     
      The corresponding numerical values 
      of the FMD-AM phase transition can be easily obtained
      from the data in the second column of table \ref{TAB1}
      by making use of the symmetry (\eq{F_TRAFO}).
\end{itemize}
\begin{table}
\begin{center}
\begin{tabular}{|| c || c |  c || c ||} \hline \hline
         & \multicolumn{2}{c||}{mean-field} \\ 
         & \multicolumn{2}{c||}{approximation} \\ \hline
$\tk$    & $\k_{{\rm FM-FMD}} $ & $\f$ \\ \hline \hline
$0.013$  & $0.01467$          & $0.21887$  \\ \hline  
$0.014$  & $0.01566$          & $0.42089$  \\ \hline  
$0.015$  & $0.01597$          & $0.52455$  \\ \hline  
$0.020$  & $0.01639$          & $0.73962$  \\ \hline  
$0.025$  & $0.01621$          & $0.81831$  \\ \hline  
$0.030$  & $0.01609$          & $0.85916$  \\ \hline  
$0.050$  & $0.01608$          & $0.92369$  \\ \hline  
$0.1$    & $0.01626$          & $0.96377$  \\ \hline  
$1$      & $0.01649$          & $0.99651$  \\ \hline  
$10$     & $0.01651$          & $0.99965$  \\ \hline  
$100$    & $0.01651$          & $0.99997$  \\ \hline  
$\infty$ & $0.01652$          & $1.00000$  \\ \hline \hline  
\end{tabular}
\end{center}
\caption{{\em The mean-field results 
for the critical coupling $\k_{{\rm FM-FMD}} (\tk)$ and the order parameter 
$\f$ evaluated at $\k=\k_{{\rm FM-FMD}}(\tk)$ are given in columns two and three 
for several values of $\tk$.
The value of $\k_{{\rm FM-FMD}}$ at $\tk = \infty$ was calculated from     
eq.~({\protect \eq{FM-FMD_IN}}). 
}}
\label{TAB1}
\end{table}
The mean-field phase diagram for $d=4$ is displayed in 
fig.~\ref{phase_diagram}. The phase boundaries are represented in this plot  
by the solid lines.  The FM-FMD phase transition line was computed by solving  
eqs.~(\eq{SPE3}), (\eq{SPE1_C}) and (\eq{SPE2_C}) numerically. The FMD-AM 
transition was computed from the the
FM-FMD phase transition data by making  
use of the symmetry (\eq{F_TRAFO}). The various symbols in 
fig.~\ref{phase_diagram} represent the results 
of the numerical simulation and will be explained later.
%
\section{Weak Coupling Expansion}
\label{PERT}
%
Eq.~(\eq{TKG2}) suggests that the weak coupling expansion should be 
performed in $1/\tk$. We consider small fluctuations 
of the $\phi$ fields around the classical 
ground state, which we parametrize by the  Goldstone field $\th_x$: 
\be
\phi_x = \exp \left( i \; \sum_\m q_\m \; x_\m + i\; \theta_x / \sqrt{2 \tk} \right)  \;.           
\lb{FLUCTUATION}
\ee
This ansatz holds both in the FM phase, where $q_\m=0$ and also 
the FMD phase, where $q_\m \neq 0,\pi$ for at least one component 
$\m=1,\ldots,4$. 
The phases $q_\m$ in the FMD phase can be computed by minimizing the 
effective potential, or alternatively could be taken from the numerical 
simulation.  

To calculate $v$ in perturbation theory, we first 
insert eq.~(\eq{FLUCTUATION}) into the action (\eq{ACTION}) of the reduced 
model and expand it in powers of $\th_x$, 
\bea
S \!\!& = &\!\! \half \int_k \; \th (-k) \; \D_q^{-1} (k) \; \th (k)  \nnn
\!\!& + &\!\! 
\int_{k_1,k_2,k_3,k_4} \;
{\cal V}_q (k_1,k_2,k_3,k_4) \; 
\; \th (k_1) \; \th (k_2)\; \th (k_3)\; \th (k_4)+\ldots \;, \lb{ACTION_THETA}
\eea
where 
\be
\int_k =\int_{0}^{2 \pi} \frac{d^4 k}{(2\pi)^4} \;.   \lb{MOM_INT}           
\ee
The subscript $q$ indicates that the propagator $\D_q (k)$ and  
four-point vertex function ${\cal V}_q (k_1,k_2,k_3,k_4)$
depend on the phases $q_\m$.     
The inverse propagator $\D_q^{-1} (k)$ is given by 
\bea
\D_q^{-1} (k) \!\!& = &\!\! \left\{ \sum_\m \cos q_\m \; 2 \; (1-\cos k_\m ) 
\right\}^2
- 8\; \left( F(q)-4 \right) \sum_\m \cos q_\m \; (1-\cos k_\m ) 
\nonumber \\
 \!\!& + &\!\! \left\{ 2 \; \sum_\m \sin q_\m \; \sin k_\m \right\}^2 
	       + 4 \; \sum_{\m \n} \sin^2 q_\m \; \sin^2 q_\n \; (1-\cos k_\n) 
\nonumber \\
 \!\!& - &\!\! \left\{ \sum_\m \sin 2q_\m \; \sin k_\m \right\}^2
	       - 2 \; \sum_{\m \n}\sin^2 q_\n \; \cos^2 q_\m \; \sin^2 k_\m    
\nonumber \\
 \!\!& + &\!\! m^2 \; \sum_\m \cos q_\m \; 2 \; (1-\cos k_\m ) \;, \lb{PROP}
\eea
with 
\be
m^2=\frac{\k}{\tk}  \;. \lb{MASS2}           
\ee
(The expression for the propagator simplifies if $q$ is a 
solution of the classical saddle-point equations, cf. eq.~(\eq{SCLSADDLEL}) below.
As we will discuss below, however, this is not in general the case in finite volume.)
At tree-level,  the FM-FMD phase transition line
is given by $m^2=0$, cf. eq. (\eq{TREE}). 
When approaching the FM-FMD phase transition line, eq.~(\eq{PROP}) reduces to      
\be
\D_{q} (k) \propto 1/(k^2 )^2 \;,     
\;\;\;\; \k \ra \k_{{\rm FM-FMD}}  \lb{PDIV}      
\ee
for small $k$. 
 The propagator (\eq{PDIV}) leads to infra-red divergences. A similar  
situation is encountered in two space-time dimensions where 
infra-red divergences occur for massless bosons with an ordinary kinetic term 
\cc{Me}.   
These infra-red divergences are not only an artifact 
of the tree-level propagator, but occur in the full theory when 
the continuum limit $\k \ra \k_{{\rm FM-FMD}}$ is performed. 
The qualitative agreement with the two-dimensional 
behavior will be demonstrated below both analytically and
numerically.  Here we note that the situation  
we encounter in the reduced model is similar to the situation 
of the XY model in 
two dimensions. The FM-FMD phase transition line behaves like 
the spin-wave phase, where critical exponents depend continuously 
on the coupling constant. Below we will show that the magnetization 
$v$ (helicoidal magnetization $v_{{\rm H}}$)
vanishes  $\propto |\k -\k_{{\rm FM-FMD}} |^{\eta(\tk)}$ 
when $\k \searrow \k_{{\rm FM-FMD}}$ ($\k \nearrow \k_{{\rm FM-FMD}}$)
and that the critical exponent $\eta(\tk)$ depends continuously on $\tk$.

It is useful to distinguish between observables which are invariant under 
the global U(1) symmetry (symmetric), like the two-point function 
$\lag \phi_x^{\dagger} \phi_y \rag$, 
and others which are not invariant (nonsymmetric), like the magnetization 
$\lag \phi_x \rag$.
For symmetric observables, the weak coupling 
expansion is infra-red finite, because all interactions involve derivatives.  
The situation is different for nonsymmetric observables, 
such as the magnetization. The real expansion parameter 
is not $1/\tk$ in this case, but $(\log m^2) /\tk$. 
This means that in order to obtain 
a nondivergent result in the limit $m^2 \ra 0$, one should               
perform a resummation of infinitely many diagrams.

\begin{figure}
\centerline{
\epsfxsize=12.0cm
\vspace*{-.5cm}
\epsfbox{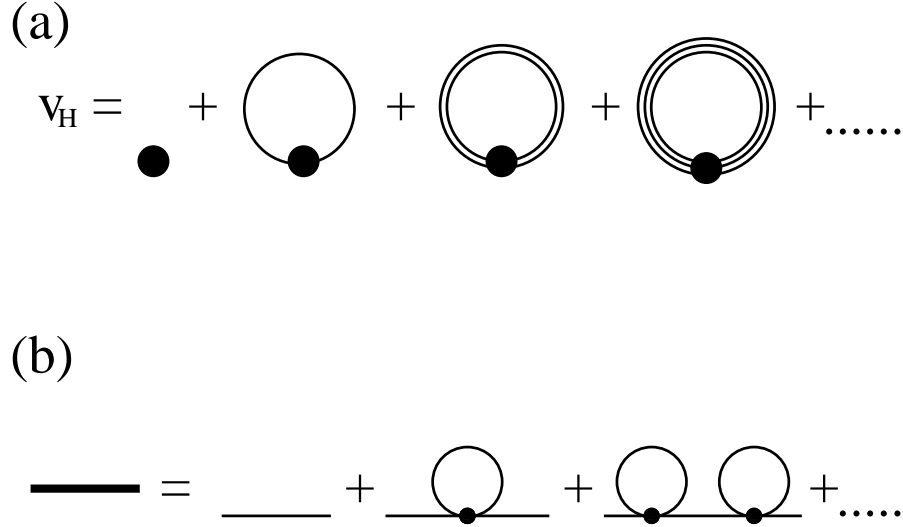}
}
\vspace*{1.2cm}
\caption{ \noindent {\em  Feynman diagrams for the magnetization (a) and     
the scalar field propagator (b).
}}
\label{FEYNMAN}
\end{figure}

Using eqs.~(\eq{FLUCTUATION}) and (\eq{ACTION_THETA}) 
the magnetization, cf. eq. 
(\eq{YDMAG}), can be calculated to one-loop in perturbation theory, 
\be
v_{{\rm H}} 
=1-\frac{1}{4\; \tk} \int_k \D_q (k) + \mbox{ higher order
corrections} \;. \lb{PMAG1}
\ee
The integral in eq.~(\eq{PMAG1}) 
is infra-red divergent in the limit $m^2 \ra 0$ and, as mentioned 
in the previous paragraph, in order to obtain a finite result we have to resum the 
higher order diagrams (with two and more lines) in fig.~\ref{FEYNMAN}a
which arise from the  terms proportional to  $\th^{2n}_x$
in eq.~(\eq{FLUCTUATION}) with $n>1$. This resummation of diagrams gives
\bea
v_{{\rm H}}
\!\!& = &\!\!
\exp \left( -\frac{1}{4\; \tk} \int_k \D_q (k)  \right) + \mbox{ higher order
corrections}
\lb{PMAG2} \\
\!\!& \sim &\!\! \left(\frac{|\k|}{\tk}\right)^{\eta } + \mbox{higher order
corrections}\;, \lb{PMAG3}
\eea
where   we used that at tree-level $\k_{{\rm FM-FMD}}=0$.
The higher order corrections in eqs.~(\eq{PMAG2}) 
and (\eq{PMAG3}) are due to quartic and higher 
order interactions which we have ignored.         
The critical exponent $\eta$ in eq.~(\eq{PMAG3}) 
is given by 
\be
\eta=\frac{1}{64 \pi^2 \tk}  \;.  \lb{EETA}
\ee
Eq.~(\eq{PMAG3}) shows that the 
magnetization vanishes at the FM-FMD phase transition with a critical 
exponent $\eta$
that depends on $\tk$ and differs from the Gaussian exponent $1/2$ of 
the XY model at $\tk=0$. (In order to show that also the helicoidal magnetization 
$v_{{\rm H}}$ vanishes for $\k \nearrow \k_{{\rm FM-FMD}}$ with the critical 
exponent (\eq{EETA}), one uses eq.~(\eq{PROP}) where $q$ is the nontrivial 
solution of eq.~(\eq{SCLSADDLEL}) below.)

Using relation (\eq{PMAG2}) and replacing the integral by a lattice sum,
we can compute $v_{{\rm H}}$ on a 
finite volume at any value of $\tk$ and $\k$ in the FM and FMD phases 
provided that we know the function $q=q(\tk,\k)$.  
On a finite lattice with periodic boundary conditions for the scalar fields
the phases $q_\m$ can only take the values           
\be
q_\m= 2 n_\m \p /L \;, \;\;\;\;\;\;n_\m=0,\ldots,L-1 \;, \lb{QDIS}
\ee
where $L$ designates 
the extent of the lattice in temporal and spatial directions. 
The $q$'s can be calculated at tree level 
by minimizing the classical action 
\be
S_0(q;\tk,\k)=L^4 \; \left\{ -2\;(16 \; \tk +\k) \; F(q) +4 \; \tk\; F(q)^2 -
\tk \; \left(\sum_\m \sin^2 q_\m \right)^2 \right\}
\;, \lb{SCLASSICAL}
\ee
where $L^4$ is the volume of the lattice and $F(q)$ is defined in 
eq.~(\eq{FQ}). We note that after expanding eq.~(\eq{SCLASSICAL}) 
in powers of $q$ and setting $q_\m= g A_\m$, we recover to leading 
order the classical potential in eq.~(\eq{VCL}). 
In infinite volume, the phases $q$ 
can be determined from the four saddle-point equations,
\be
\left[ 2\; (16 \; \tk +\k) -8\; \tk \; F(q) - 4 \; \tk \; 
\left( \sum_\m \sin^2 q_\m \right) \;
\cos q_\n \right]  \; \sin q_\n =0 \;.
 \lb{SCLSADDLEL}
\ee
In finite volume, the value of $q$ that minimizes $S_0(q;\tk,\k)$
at a given $(\tk, \k)$ point in the FMD phase
is determined by computing $S_0(q;\tk,\k)$ for all $L^4$ vectors $q$ 
numerically, 
and picking out the ones with the smallest value of $S_0(q;\tk,\k)$.           
(The minimum will respect the lattice symmetries and hence  
will in general not be unique.) {}From the resulting 
$q$-values we have computed the observable 
$F(q)$, cf. eq.~(\eq{FQ}), for several $(\tk,\k)$ points in the FMD phase. 
The discussion of this quantity will be postponed to 
Sect.~\ref{CPERT} where we will compare it with the results of the 
numerical simulation.

We will from now on focus on 
the physics in the FM phase where $q=0$,
and calculate  the magnetization to one higher order in 
perturbation theory. {}From this calculation we will obtain 
another estimate for the critical coupling $\k_{{\rm FM-FMD}}(\tk )$. 

The vertex function ${\cal V}_0 (k_1,k_2,k_3,k_4)$ in eq.~(\eq{ACTION_THETA})
is given by    
\bea
&&\!\!\!\!\!\!\!\! {\cal V}_0 (k_1,k_2,k_3,k_4) = -
\frac{1}{4 \; \tk} \left\{ - \left[ \sum_\m (1-\cos (k_1+k_2)_\m ) \right]^2 
               +\fthird \; \left[ \sum_\m (1-\cos {k_1}_\m )  \right]^2  \right. \nnn 
&&\!\!\!\!\!\!\!\! +\sum_{\m \n} \sin {k_1}_\m \; \sin {k_2}_\m \;
			   \sin {k_3}_\n \; \sin {k_4}_\n 
              + \half \; m^2 \; \sum_\m (1-\cos (k_1+k_2)_\m ) \nnn
&&\!\!\!\!\!\!\!\! -\left. \twothird \; m^2 \; \sum_\m (1-\cos {k_1}_\m )  \right\} \; 
	     \d (k_1+k_2+k_3+k_4)  \;, \lb{VERTEX}        
\eea
where the first three terms arise from the 
$\tk \sum_x \{ \phi^{\dagger}_x (\Box^2 \phi)_x - B_x^2 \}$ term and the 
two terms proportional $m^2$ from the 
$-\k \sum_x \phi^{\dagger}_x (\Box \phi)_x$ term in the action (\eq{ACTION}). 
We note that the vertex function (\ref{VERTEX}) can be easily rewritten such 
that it is symmetric with respect to the momenta $k_1$, $k_2$, $k_3$ and $k_4$. 
For the perturbative calculation it does not matter which form is used.

After carrying out the higher-order calculation, the magnetization can be written 
in the form 
\bea
v  \!\!& = &\!\!
\exp \left( -\frac{1}{4\; \tk} \int_k \D_{0,{\rm 1-loop}}(k) \right)
+ \mbox{ higher order corrections} \;, \lb{PMAG4} \\
\D_{0,{\rm 1-loop}} (k) \!\!& = &\!\! \frac{\D_0 (k)}{  1+ \D_0 (k) \S (k) }
=\frac{1}{ \D_0^{-1} (k) + \S (k) }
\lb{PMAG5}
\eea
where $\D_0 (k)$ is the tree-level propagator in the FM phase (cf. eq. (\eq{PROP}) 
with $q=0$)
and $\S (k)$ is the self-energy, 
\bea
\S(k) \!\!& = &\!\!
\frac{1}{\tk} \; \int_p \; \left\{ 2\; \left[ \sum_\m (1-\cos (p+k)_\m ) \right]^2 
	       \right. \nnn
\!\!& - &\!\!2\; \left( \left[ \sum_\m (1-\cos {k}_\m )  \right]^2 +
			   \left[ \sum_\m (1-\cos {p}_\m )  \right]^2 \right)
	        \nnn 
\!\!& - &\!\! \sum_\m \sin^2 k_\m \; \sum_\n \sin^2 p_\n 
	  -2\; \left( \sum_\m \sin k_\m \; \sin p_\m \right)^2 \nnn    
\!\!& + &\!\! \left. m^2 \; \left( \sum_\m (1-\cos (p+k)_\m ) 
	 - \sum_\m (1-\cos {k}_\m ) - \sum_\m (1-\cos {p}_\m ) \right) \right\} \;
	     \D_0(p)               \;.  \nnn
\!\!&  &\!\! 
	     \lb{SELFE}        
\eea
The propagator (\eq{PMAG5}) already involves a resummation of diagrams shown in
fig.~\ref{FEYNMAN}b. The magnetization in eq.~(\eq{PMAG4})  has been 
obtained by performing the resummation of diagrams 
in fig.~\ref{FEYNMAN}a, but now using the propagator $\D_{0,{\rm 1-loop}} (k)$
instead of $\D_q (k)$. 
In order to compare the perturbative formulas (\eq{PMAG2})
and (\eq{PMAG4}) with the results of the 
numerical simulations (see Sect.~\ref{CPERT}) we have  to evaluate 
the lattice integrals in eqs.~(\eq{PMAG2}) and (\eq{PMAG4}) on a finite 
lattice. The integrals are replaced by sums over the lattice momenta.
In this context we note that these 
finite lattice sums do not include the zero mode, $k=0$. The zero mode 
decouples  from the action, and gives rise to a phase which 
disappears after taking the modulus in the definition 
of $v$ and $v_{{\rm H}}$ in eqs.~(\eq{YMAG}) and (\eq{YDMAG}).

The critical coupling can be calculated by expanding 
$\D_{0,{\rm 1-loop}} (k)^{-1}$ for small momenta in powers of $k^2$,  
\be
\D_{0,{\rm 1-loop}} (k)^{-1} =  \D_0 (k)^{-1} + \S (k) 
= a(\tk,\k) \; (k^2) + b(\tk,\k) \; (k^2)^2 +\ldots
\lb{EXPAN}
\ee
and equating the coefficient $a(\tk,\k)$ to zero,
\bea
a(\tk,\k)\!\!& = &\!\! \frac{\k}{\tk}+\frac{1}{\tk} \; \int_k 
                  \left\{ \half \; \sum_\m (1-\cos {k}_\m ) \; \sum_\n \cos {k}_\n 
                - \sum_\m \sin^2 k_\m \right. \nnn
         \!\!& + &\!\! \left. \frac{\k}{\tk} \; \left( 
	 \eigth \;\sum_\m \cos^2 k_\m -\half  \right) \right\} 
	 \; \D_0 (k) =0 \;.
\lb{XKAPPA_C}
\eea
This leads to the one-loop estimate 
\bea
\k_{{\rm  FM-FMD}} (\tk) \!\!& = &\!\! -\int_k \left\{                  
 \half \; \sum_\m (1-\cos {k}_\m ) \; \sum_\n \cos {k}_\n 
 - \sum_\m \sin^2 k_\m \right\} \nnn  
\!\!& \times &\!\! \left[ \sum_\m (1-\cos k_\m ) \right]^{-2} \lb{KAPPA_C} \\
\!\!& \approx &\!\! 0.02993 \;,
\lb{V_KAPPA_C}
\eea
which is about a factor two larger than the mean-field value, cf.~table \ref{TAB1}.
\section{Numerical Results}
\label{NUM}
\subsection{Phase Diagram}
\label{NUM_PD}
 To simulate the reduced model defined by the path integral (\eq{PATH}) 
we have implemented two different Monte Carlo algorithms, 
a five-hit Metropolis and a Hybrid Monte Carlo 
algorithm. The results for the various observables 
agree nicely within the AM, FM and PM phases. We find however that the 
Hybrid Monte Carlo algorithm gets much more easily stuck 
in metastable non-equilibrium states in the FMD phase. 
We therefore have generated the bulk 
of the data presented in this paper with a five-hit Metropolis algorithm. 

To map the phase diagram we have measured the following observables:
\begin{itemize}
\item The {\em magnetization} 
\be
v=\left\lag \left| \frac{1}{L^4} \sum_x \phi_x \right| \right\rag \;,    \lb{MAG}
\ee
which is the order parameter for ferromagnetism and  
\item the {\em staggered magnetization}
\be
v_{{\rm AM}}=\left\lag \left| \frac{1}{L^4} \sum_x \phi_x \; \e_x \right| \right\rag \;,    \lb{STMAG}
\ee
which is the order parameter for anti-ferromagnetism. 
\item The {\em helicoidal magnetization}
\be
v_{{\rm H}}=\left\lag \left| \frac{1}{L^4} \sum_x \phi_x \; \exp \left( -i \sum_\m q_\m x_\m \right) \right| \right\rag
  \lb{DMAG}
\ee
was used to map the FMD phase, where the four 
real phases $q_\m$, $\m=1,\ldots,4$
were determined for each configuration from
\be
q_\m = \mbox{Im} \; \mbox{Log} \; \left[ \frac{1}{L^4} \sum_x \phi_x^{\dagger} \phi_{x+\hmu} \right] 
\;.  \lb{QU}
\ee
\item Apart from these quantities we have also measured the internal energy
density 
\be
z^2=\left\lag \frac{1}{4 L^4} \sum_{\m x} 
\mbox{Re} \; \left(\phi_x^{\dagger} \phi_{x+\hmu} \right) \right\rag    
  \lb{Z2}
\ee
of the mass counterterm and 
\item the quantity 
\be
c(q_\m)=\lag \cos q_\m \rag \;, \lb{CMU}
\ee
where the phases $q_\m$ were calculated for each configuration 
by means eq.~(\eq{QU}). 
\end{itemize}
\begin{figure}
\vspace*{-0.5cm}
\begin{tabular}{ll}
 \hspace*{-2.3cm} \epsfxsize=10.40cm
 \vspace*{-0.5cm}
\epsfbox{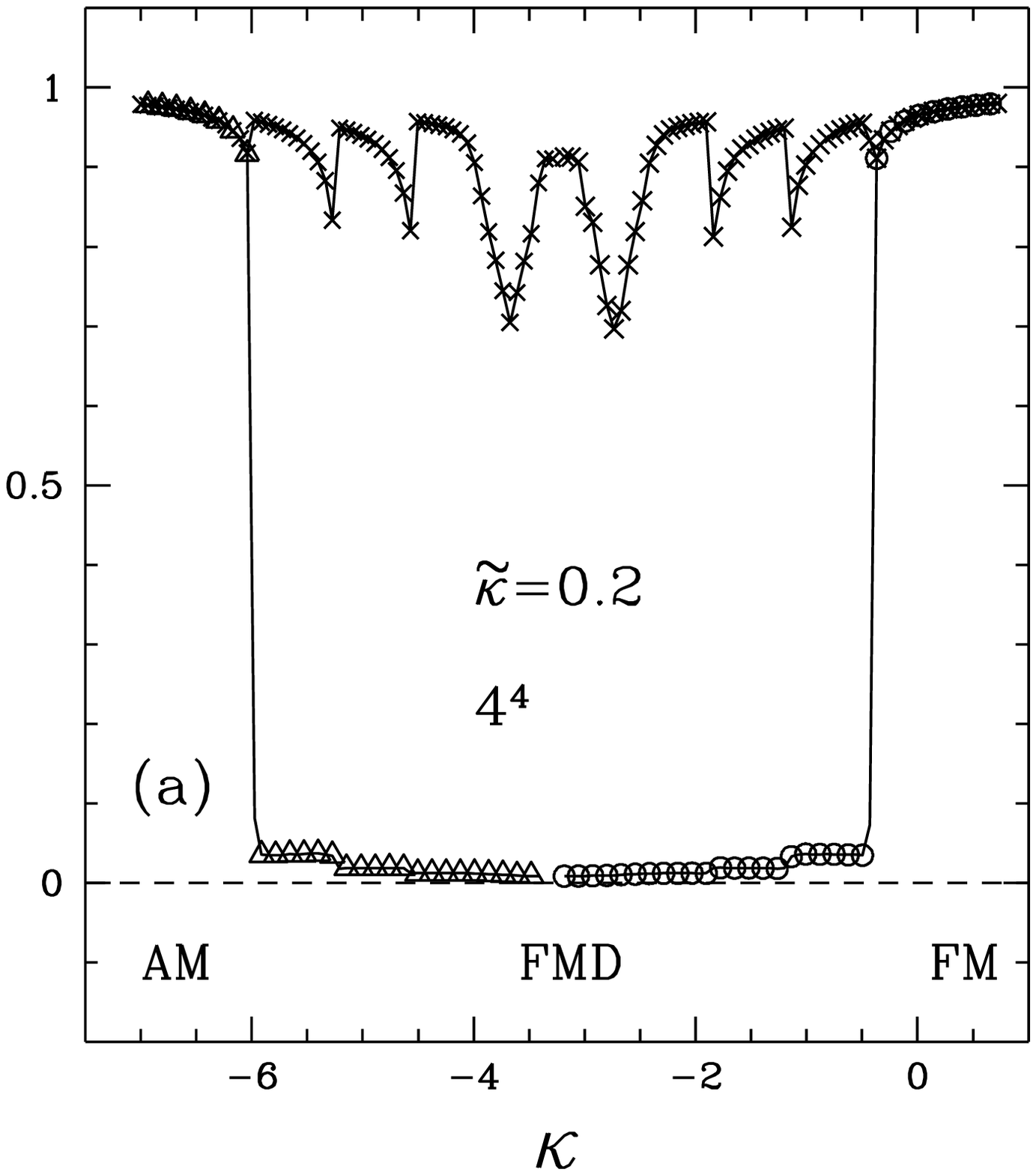}     &
 \hspace{-4.2cm} \epsfxsize=10.40cm
 \vspace*{-1.0cm}
\epsfbox{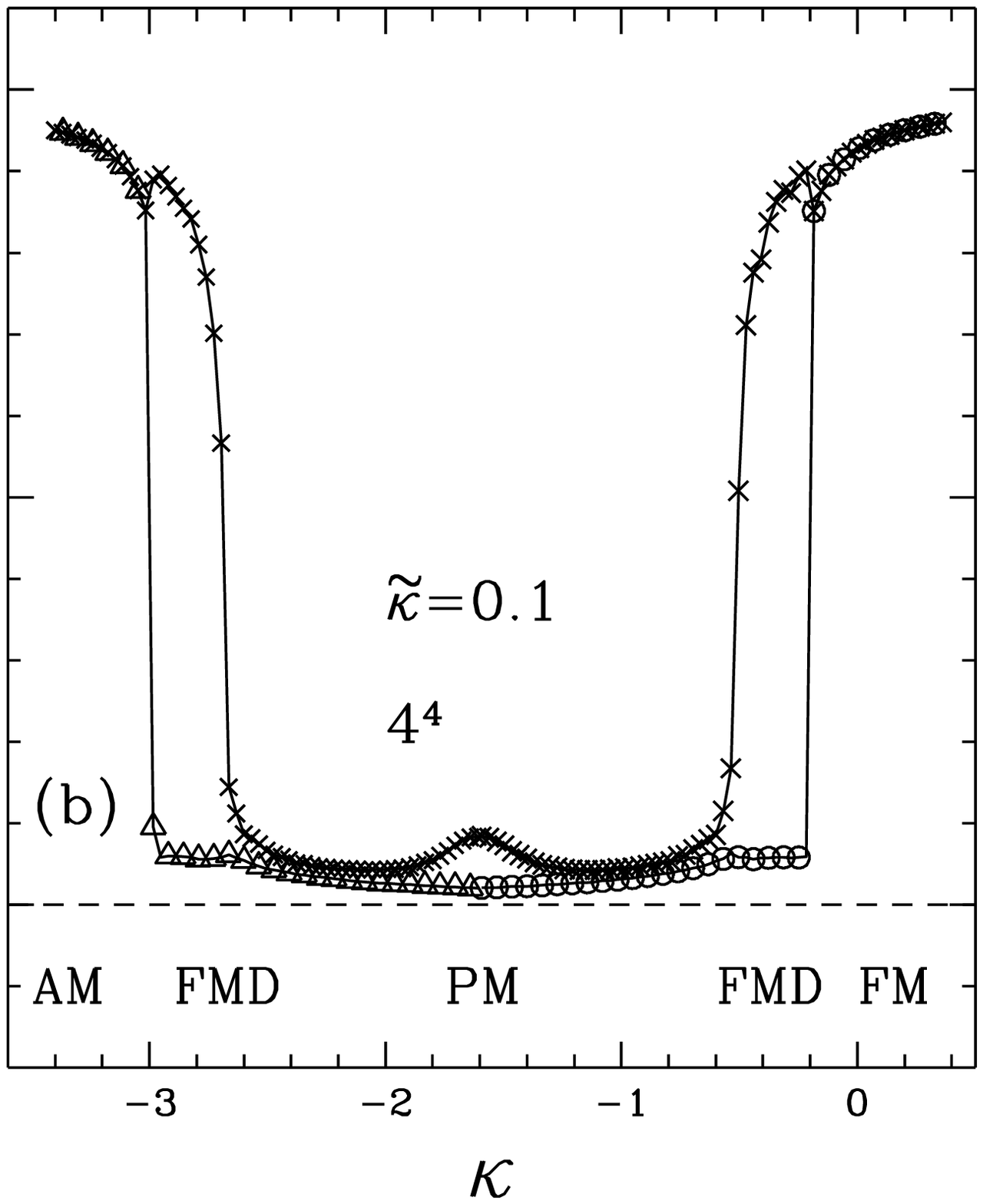}     \\
 \hspace*{-2.3cm} \epsfxsize=10.40cm
 \vspace*{-0.5cm}
\epsfbox{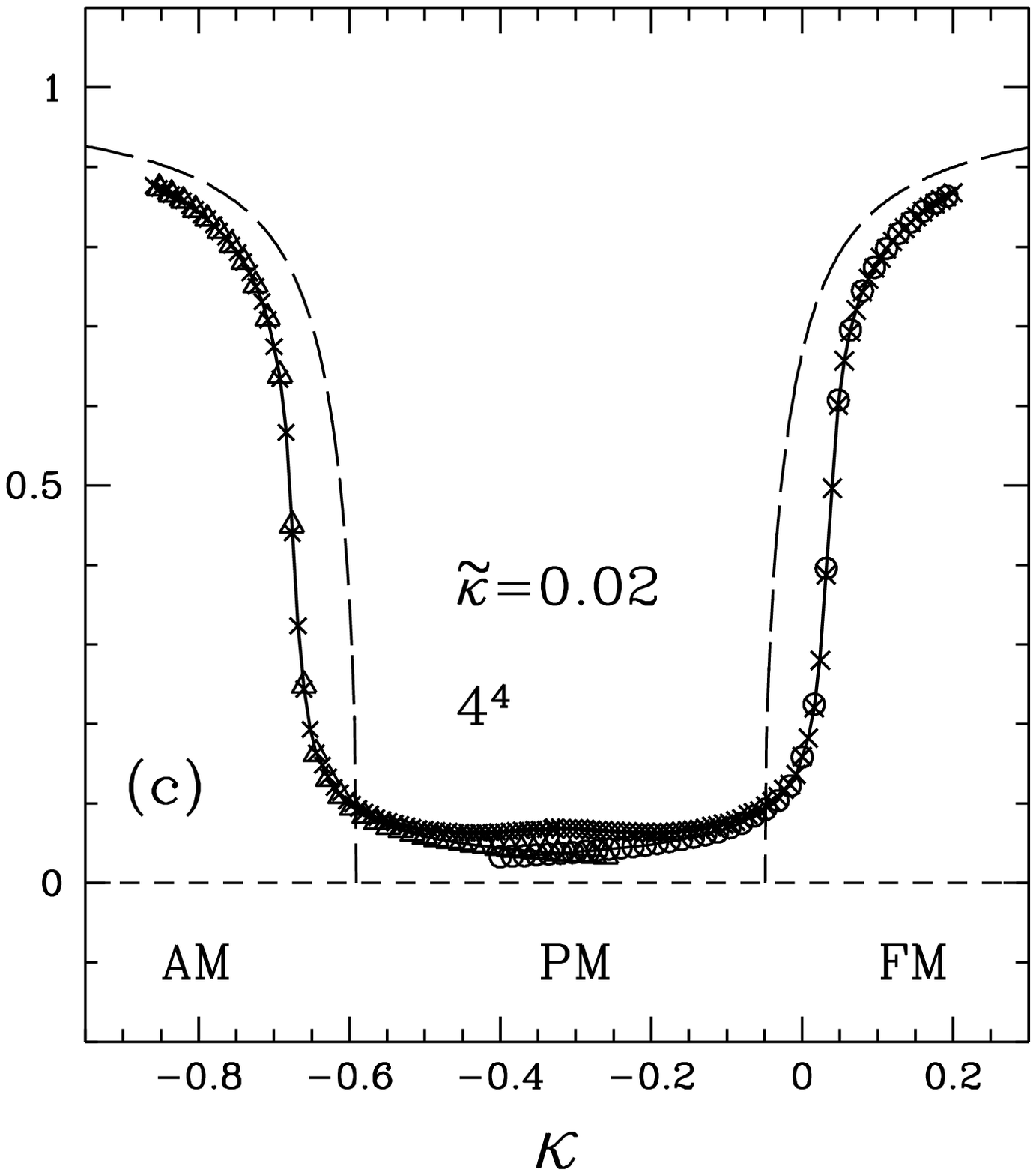}     &
 \hspace{-4.2cm} \epsfxsize=10.40cm
 \vspace*{-1.0cm}
\epsfbox{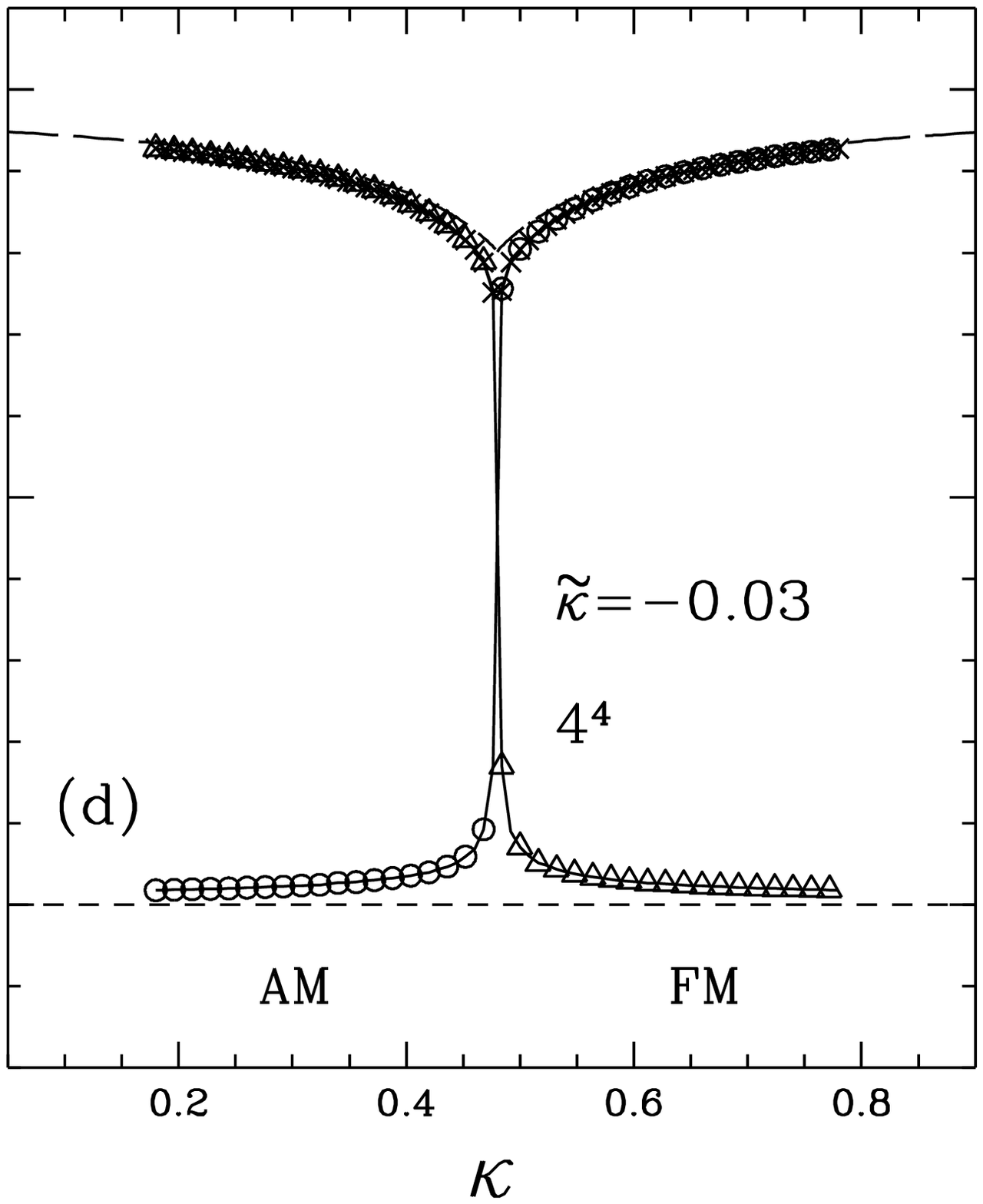}     \\
\end{tabular}
\vspace*{-0.0cm}
\caption{ \noindent {\em  Scans in $\k$ direction on 
a $4^4$ lattice at $\tk=0.2$ (fig.~a) $\tk=0.1$ (fig.~b)
$\tk=0.02$ (fig.~c) and $\tk=-0.03$ (fig.~d). 
The results for $v$, $v_{{\rm AM}}$ and $v_{{\rm H}}$ 
are represented in the plots by the circles, triangles and crosses.
The error bars are omitted because they are in all cases smaller than the 
symbol size. The dashed lines in figs.~c and d 
are obtained in the mean-field approximation.
}}
\label{SCANS}
\end{figure}
%
%
 We have taken the modulus in eqs.~(\eq{MAG}-\eq{DMAG}) for each configuration,
because, in a finite volume, the constant field mode gives rise 
to a slow drift of the magnetization through the group space. 
 (Taking the  absolute value is a standard method which allows us to avoid 
the introduction of an external magnetic field.)

The discreteness 
of the phases (\eq{QDIS}) poses a problem for the simulation in the FMD phase
because  each transition of one $q$ to another $q$ behaves very much like 
a first order phase transition and hence is accompanied by metastabilities.
We find that these metastabilities 
become more severe when the lattice size is increased. 

To determine the phase diagram we kept the parameter 
$\tk$ fixed and performed simulations at a large number of 
$\k$ values.  Each of these 
vertical scans, cf.~fig.~\ref{phase_diagram}, has been started in the 
FM phase. We lowered $\k$ in fixed steps 
and used the last configuration of a run as the initial configuration 
at the next smaller value of $\k$. At each point we skipped $10^3$ sweeps 
for equilibration and performed $10^4$ measurement sweeps. 
The error of an observable $O$ was determined using the relation 
\be
\D \lag O \rag= \D \lag O \rag_{{\rm st}} \sqrt{2 \t_{{\rm int}}} \;, \lb{ER}
\ee
where $\D \lag O \rag_{{\rm st}}$ is the standard deviation             
and $\t_{{\rm int}}$ designates
the integrated autocorrelation time, defined as 
$\t_{{\rm int}}=\sum_{\D t} \G (\D t) / \G (0) $ (see for 
example \cc{So91}). The quantity    
$\G (\D t)=\lag O(t) O(t+ \D t ) \rag -\lag O \rag^2 $
is the autocorrelation function.                                             
\begin{figure}
 \vspace*{-1.0cm}
\begin{tabular}{c}
 \hspace*{1.1cm} \epsfxsize=11.30cm
\epsfbox{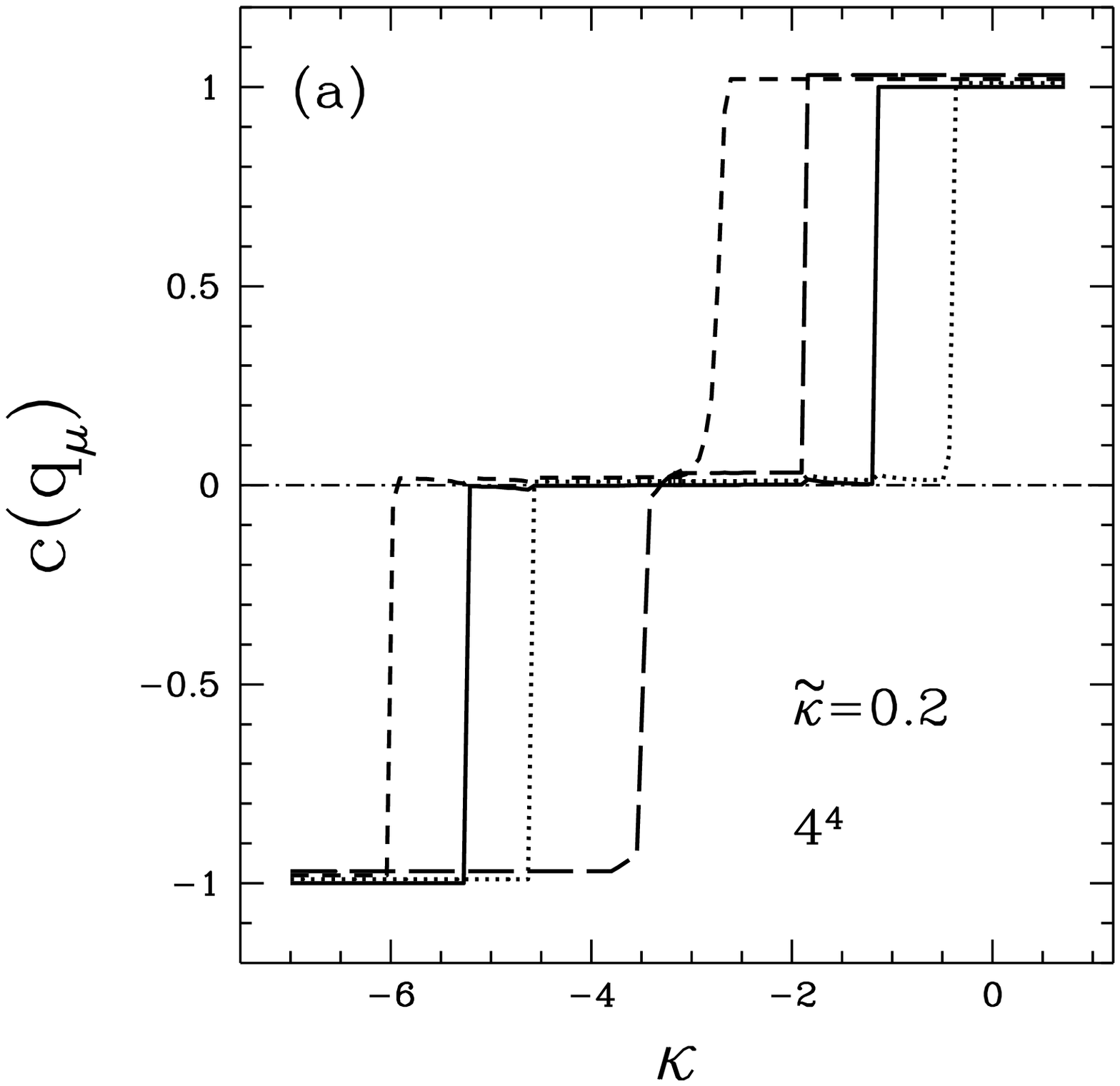}
 \vspace*{-2.0cm} \\
 \hspace*{1.1cm} \epsfxsize=11.30cm
 \vspace*{-1.4cm}
\epsfbox{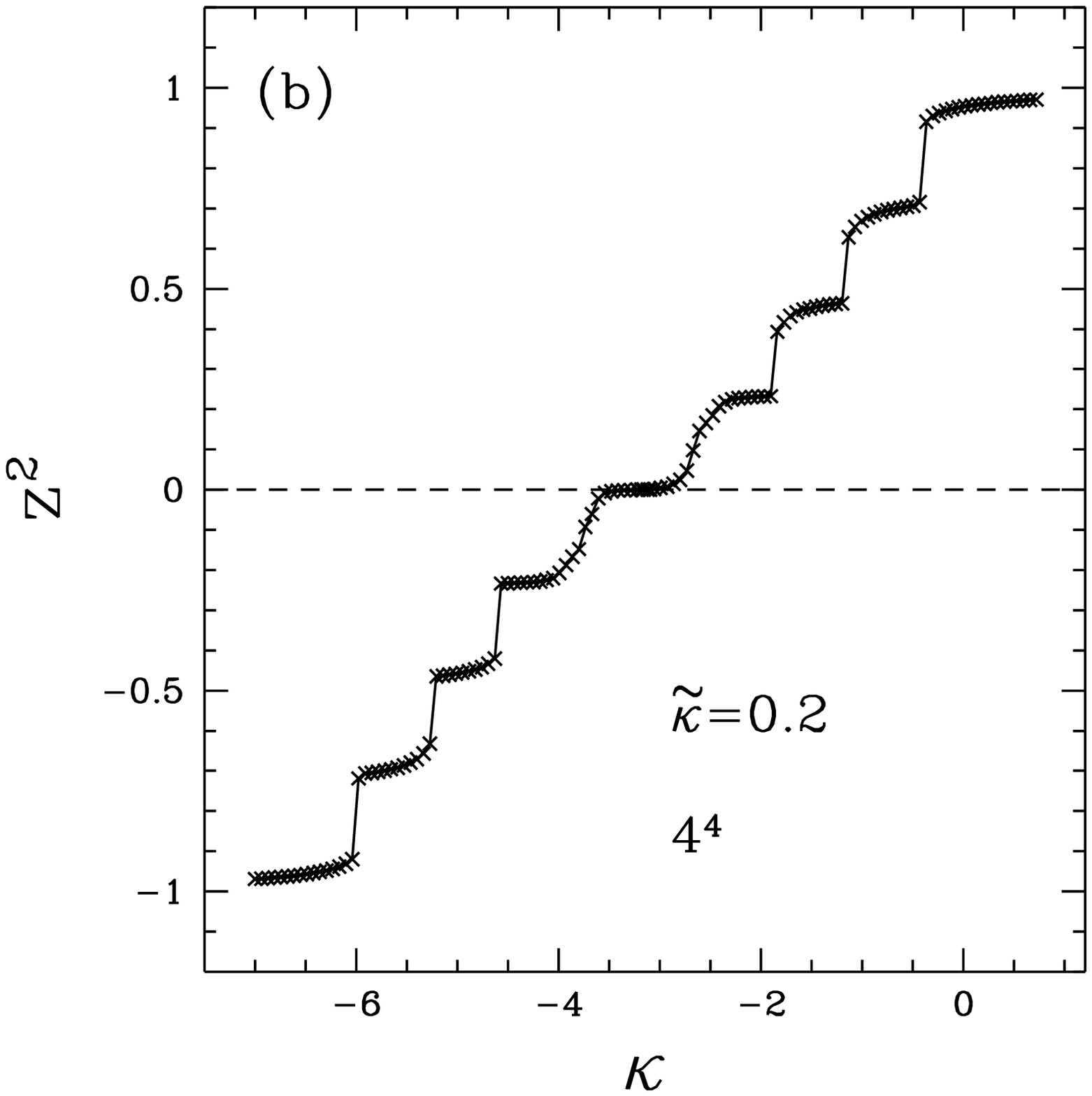}     \\
\end{tabular}
 %
\caption{ \noindent {\em  The observables $c(q_\m)$ (fig.~a) and 
$z^2$ (fig.~b) as a function of $\k$ for $\tk=0.2$. The lattice size is $4^4$.
The four observables $c(q_\m)$, $\m=1,\ldots,4$ are 
distinguished by the line type (solid, dots, long and short dashes).
The error bars are omitted in both figures because they are not larger 
than the symbol size in fig.~b and also not much bigger 
than the line width in fig.~a.
}}
\label{QUSZ2}
\end{figure}

In figs.~\ref{SCANS}a-d we have displayed the numerical results 
for the order parameters $v$ (circles), $v_{{\rm AM}}$ (triangles) and $v_{{\rm H}}$ 
(crosses) for four exemplary scans on a $4^4$ lattice.  
Fig.~\ref{SCANS}a shows that there are 
three different phases at $\tk=0.2$, an FM phase at $\k \apgt -0.43$, where 
$v=v_{{\rm H}} >0$, $v_{{\rm AM}}=0$, an AM phase at $\k \aplt -5.97$, where 
$v_{{\rm AM}}=v_{{\rm H}} >0$, $v=0$ and the FMD phase at intermediate $\k$ in which 
$v_{{\rm H}}>0$, $v=v_{{\rm AM}}=0$. The   helicoidal
magnetization $v_{{\rm H}}$ exhibits jumps at certain  
$\k$ values within the FMD phase. These jumps in $v_{{\rm H}}$ occur
because the phases $q_\m$ can change only in discrete steps and 
hence have to be considered as a finite volume artifact. 
In Fig.~\ref{QUSZ2}a we have plotted  the quantity
$c(q_\m)$, $\m=1,\ldots,4$ for the same scan at $\tk=0.2$ 
as a function of $\k$. A comparison of 
figs.~\ref{SCANS}a and \ref{QUSZ2}a shows that the jumps in $v_{{\rm H}}$ occur at 
the same $\k$ values where one of the components $c(q_\m)$ 
exhibits a jump. All $q_\m$'s are zero in the FM phase. 
At $\k \approx -0.43$ the first component of $q$ condenses (dotted line) and becomes 
equal to $2\p/L=\p/2$. The next jump occurs when also the second component of $q$ 
becomes equal to $\pi/2$ (solid line). The $\k$ value where finally all values of 
$q$ are equal to $\p/2$ coincides nicely with the 
symmetry point, $\k=-16 \tk = -3.2$, cf. eq.~(\eq{F_TRAFO}). 
We note that the order in which the jumps occur
is arbitrary (because of hypercubic symmetry). The 
jumps at  $\k< -16 \tk$ follow a similar pattern, with 
the $q_\m$'s jumping from $\p /2 $ to $\p$. 
We will show in Sect.~\ref{CPERT} that the complicated $\k$-dependence  
of $v_{{\rm H}}$ in the FMD phase can at least  qualitatively  be explained
by the one-loop formula (\eq{PMAG2}). In fig.~\ref{QUSZ2}b we have plotted 
the internal energy density $z^2$ as a function of $\k$. 
In the mean-field theory this quantity is given 
by  $v_{{\rm H}}^2 \quart \sum_\m \cos q_\m$ and since  $v_{{\rm H}}^2=O(1)$, 
we expect this quantity to jump whenever $q$ changes.
The comparison of figs.~\ref{QUSZ2}a and \ref{QUSZ2}b shows that this 
is indeed the case.

A different situation is encountered in fig.~\ref{SCANS}b which shows 
the result of the scan at $\tk=0.1$. Besides the FM, AM and FMD phases we find 
now clear evidence for a PM phase (where $v=v_{{\rm AM}}=v_{{\rm H}}=0$) which, 
as predicted by the mean-field calculation, extends into the FMD phase. The graph 
shows that four different phase boundaries are crossed when $\k$ is lowered 
from the FM to the AM phase. 
The small peak at $\k=-16 \tk \approx -1.6$ appears to be a finite size effect
because it becomes smaller when the lattice size is enlarged.

The FMD phase gradually disappears when $\tk$ is lowered further.
The situation at $\tk=0.02$ is depicted in fig.~\ref{SCANS}c.
The FMD phase has now completely disappeared, and the only three phases we are 
left with are the FM, PM and AM phases. We 
find that the PM phase extends down to $\tk \approx -0.02$. 
The result of the scan at 
$\tk =-0.03 $ is displayed in fig.~\ref{SCANS}d. It shows that the phase transition 
between the FM and AM phase coincides with the symmetry line $\k=-16 \tk$
and is obviously of first order. Both the internal energy density $z^2$  and the 
order parameters $v$ and $v_{{\rm AM}}$ exhibit a gap at this 
phase transition. This gap grows from zero 
to one when one follows the symmetry line $\k=-16\tk$ 
from the triple point where the 
FM, PM and AM phases meet to $\tk=-\infty$. The two dashed 
lines in figs.~\ref{SCANS}c
and d represent the mean-field result for the magnetization $\f=v$
in the FM and for the staggered magnetization $\f=v_{{\rm AM}}$
in the AM phase, which we obtained by solving the mean-field equations 
(\eq{SPE1}) and (\eq{SPE3}) for $q=(0,0,0,0)$  and $q=(\p,\p,\p,\p)$ numerically. 

We have read off the positions of the various phase transitions 
from plots like the ones depicted in figs.~\ref{SCANS}a-d and 
then compiled them 
in the $\k$-$\tk$ phase diagram plot in fig.~\ref{phase_diagram}. 
The triangles were obtained on a $4^4$ lattice, 
the crosses represent the phase boundaries 
on a $6^4$ lattice and the circles mark the phase transitions points on  an $8^4$ 
lattice. The solid curves are the mean-field results which we discussed 
in Sect.~\ref{CMFA}.            
The plot shows that the numerical estimates 
for the FM-AM, FM-PM, PM-AM, FM-FMD and FMD-AM phase boundaries agree nicely 
with the mean-field prediction. The agreement seems to be                   
worse for the FMD-PM phase transition.
The numerical data indicate however that both the horizontal and vertical width 
of the PM phase 
shrink when the lattice size is increased and that  the numerical 
results could come out closer to the mean-field result for larger  volumes. 
\begin{figure}
\centerline{
\epsfysize=14.0cm
\vspace*{0.5cm}
\epsfbox{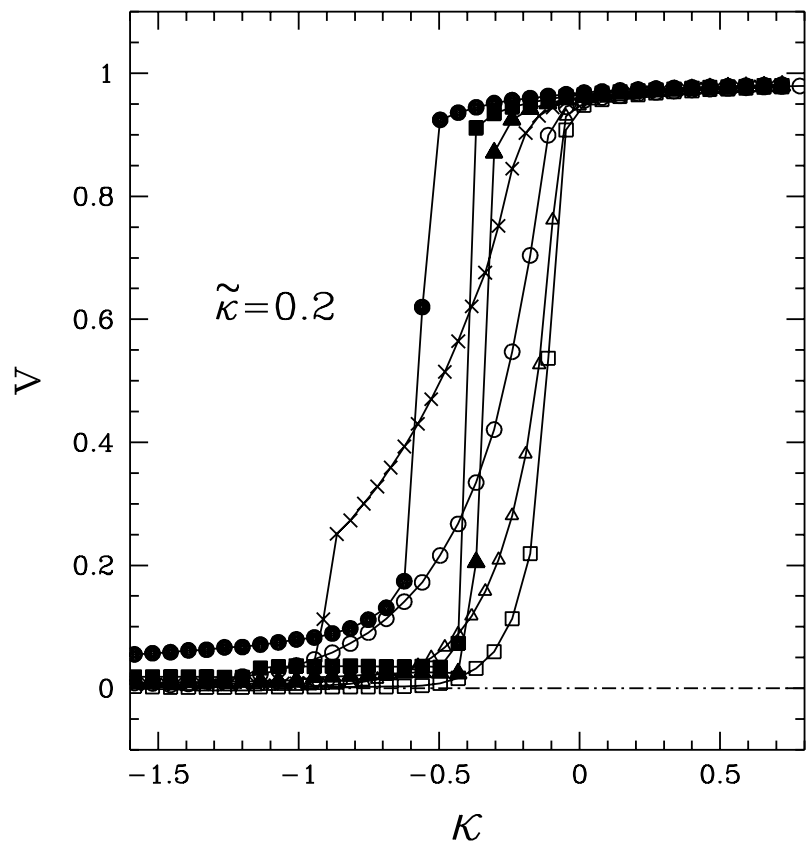}
}
\vspace*{-0.8cm}
\caption{ \noindent {\em  
The magnetization $v$ as a function of $\k$ at  $\tk=0.2$ on $3^4$ (filled circle),
$4^4$ (filled square), $5^4$ (filled triangle), $6^4$ (crosses),
$8^4$ (open circle), $10^4$ (open triangle) and $12^4$ (open square) lattices.
}}
\label{v_fs}
\end{figure}

In fig.~\ref{v_fs} we have 
displayed the magnetization $v$ as a function of $\k$ for $\tk=0.2$ and 
a series of different lattice sizes.  Again, we have lowered $\k$ in small steps, 
and used the last configuration of a run as initial configuration for the next run, 
skipping $10^3$ sweeps for equilibration.

First we discuss the results 
on the smaller lattices  of size  $3^4$  (filled circle), 
$4^4$ (filled square) and $5^4$ (filled triangles).
The magnetization $v$ exhibits 
a jump on  
these lattices. The helicoidal magnetization $v_{{\rm H}}$ is 
identical with $v$ in the FM phase and (unlike $v$) remains of $O(1)$ 
when crossing the FM-FMD phase transition towards the FMD phase 
(the data for $v_{{\rm H}}$ are not included 
in fig.~\ref{v_fs}). It can be seen that the curves for $v$ bend in more strongly 
when the lattice size is increased, indicating that $v$ scales to zero 
at the FM-FMD phase transition as predicted 
by the weak coupling expansion in Sect.~\ref{PERT}. We will demonstrate in the next 
subsection that the data for $v$ in the FM 
phase are nicely consistent  with the perturbative formula (\eq{PMAG4}) 
according to which $v \searrow 0 $ in the limit $\k \searrow  \k_{{\rm FM-FMD}}$ 
and $L \ra \infty$. The plot also shows  
that $\k_{{\rm FM-FMD}}$ increases with increasing lattice size. 

On the larger lattices we encounter a different behavior. 
The magnetization first starts to bend over when $\k$ is lowered
but then instead of jumping to 
the FMD phase continues to decrease slowly. The jump to the FMD 
occurs finally at a large negative value of $\k$. The jump 
on the $6^4$ lattice (crosses) for instance occurs at $\k\approx -0.9$ and not 
at $\k \approx -0.2$. The reason for this 
effect  could be related to the fact that the FM-FMD phase 
transition behaves very much like a first order phase transition
as the magnetization drops to zero very rapidly at $\k_{{\rm FM-FMD}}$.
On the larger lattices the fluctuations in the internal energy 
are very small and hence the probability for a jump across a large energy barrier 
becomes strongly suppressed. This situation did not change 
after increasing the statistics by one order 
of magnitude which means that the probability for a transition to occur 
at larger $\k$ is very small. 
In this context we also note that on larger lattices 
the system in the FMD phase ends up in different states when using      
different starting configurations and transitions 
to other states occur very rarely or not at all.  The 
results for the various observables in the FMD phase are 
independent of the initial configuration 
only on  the smaller lattices of size $3^4$, $4^4$ and $5^4$. (In all 
other phases, our results are independent of the initial configurations on all 
volumes.)

Fig.~\ref{v_fs} shows that the region 
in the FM phase where the magnetization starts to bend over is shifted 
in all cases towards larger values of $\k$ 
when the lattice size is enlarged. We will show in the 
next subsection that this finite size behavior is in nice agreement with 
the perturbative formula given in Sect.~\ref{PERT}. This good agreement 
between the numerical data and lattice perturbation theory lead us to 
identify $\k_{{\rm FM-FMD}}$ on the larger lattices, i.e. for
$L \geq 6$, with the point where the slope in $v$ is largest, and 
not with the point at large negative $\k$, where $v$ exhibits the jump 
and $v_{{\rm H}}$ becomes different from $v$. 
All phase transition points on the $6^4$ and $8^4$ lattice 
which are included in fig.~\ref{phase_diagram} were obtained with  
this criterion. 
\begin{figure}
\centerline{
\epsfysize=14.0cm
\vspace*{0.5cm}
\epsfbox{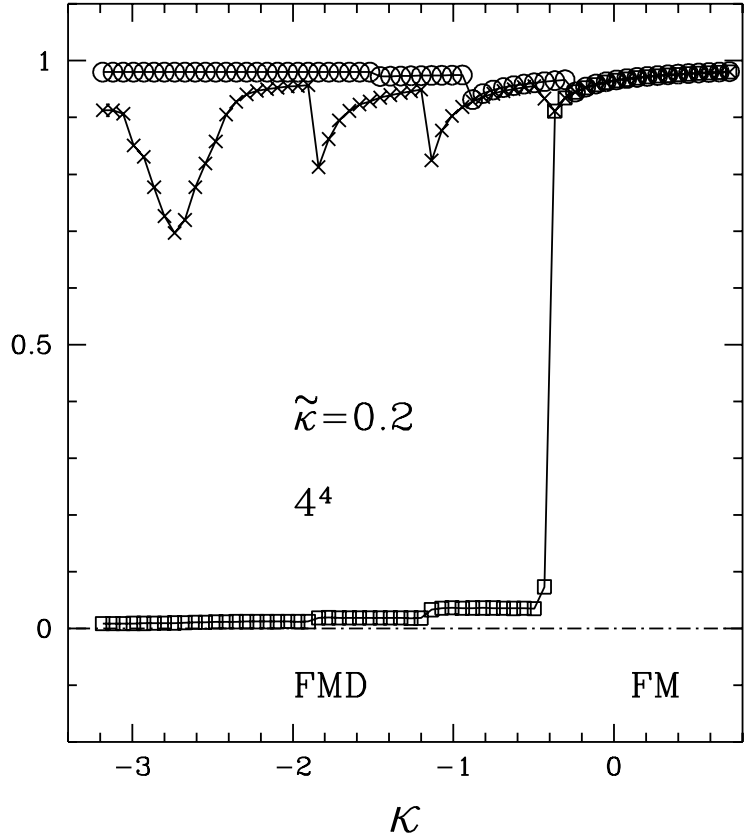}
}
\vspace*{-1.0cm}
\caption{ \noindent {\em  
The magnetization $v$ (squares) 
and the helicoidal magnetization $v_{{\rm H}}$ (crosses) 
in the FM and FMD phases as a function of $\k$ 
for $\tk=0.2$. The same data were presented already in 
fig.~{\protect \ref{SCANS}}a. The circles were obtained by numerically 
evaluating the perturbative expression ({\protect \eq{PMAG2}}). The   
phases $q_\m=q_\m (\tk,\k)$ were obtained by minimizing eq.~(\eq{SCLASSICAL}).
}}
\label{TFMD}
\end{figure}
%
%
\subsection{Comparison with Perturbation Theory}
\label{CPERT}
\begin{figure}
 \vspace*{-1.0cm}
\begin{tabular}{c}
 \hspace*{1.1cm} \epsfxsize=11.30cm
\epsfbox{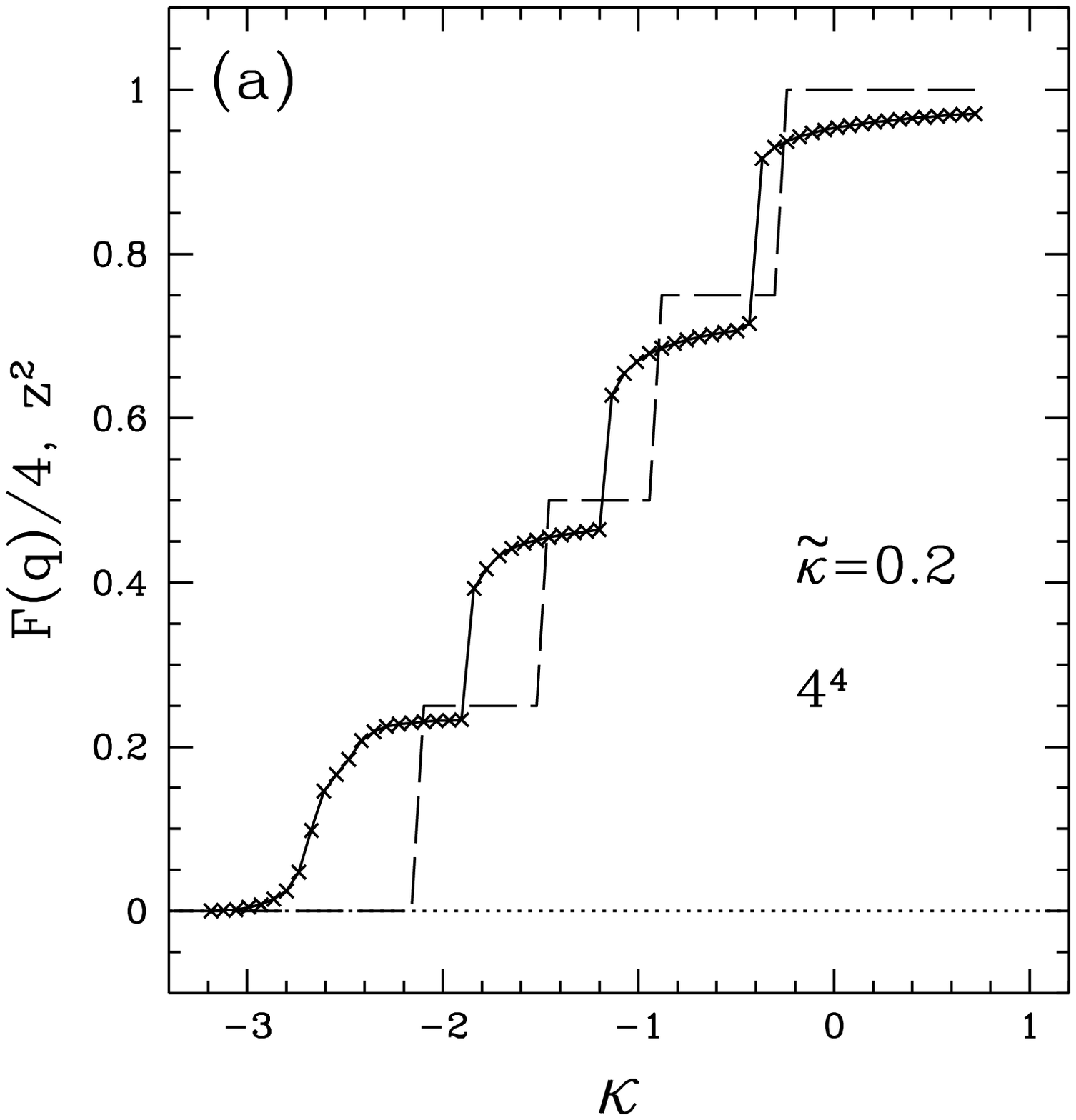}
  \vspace*{-2.0cm} \\ 
 \hspace*{1.1cm} \epsfxsize=11.30cm
 \vspace*{-1.2cm}
\epsfbox{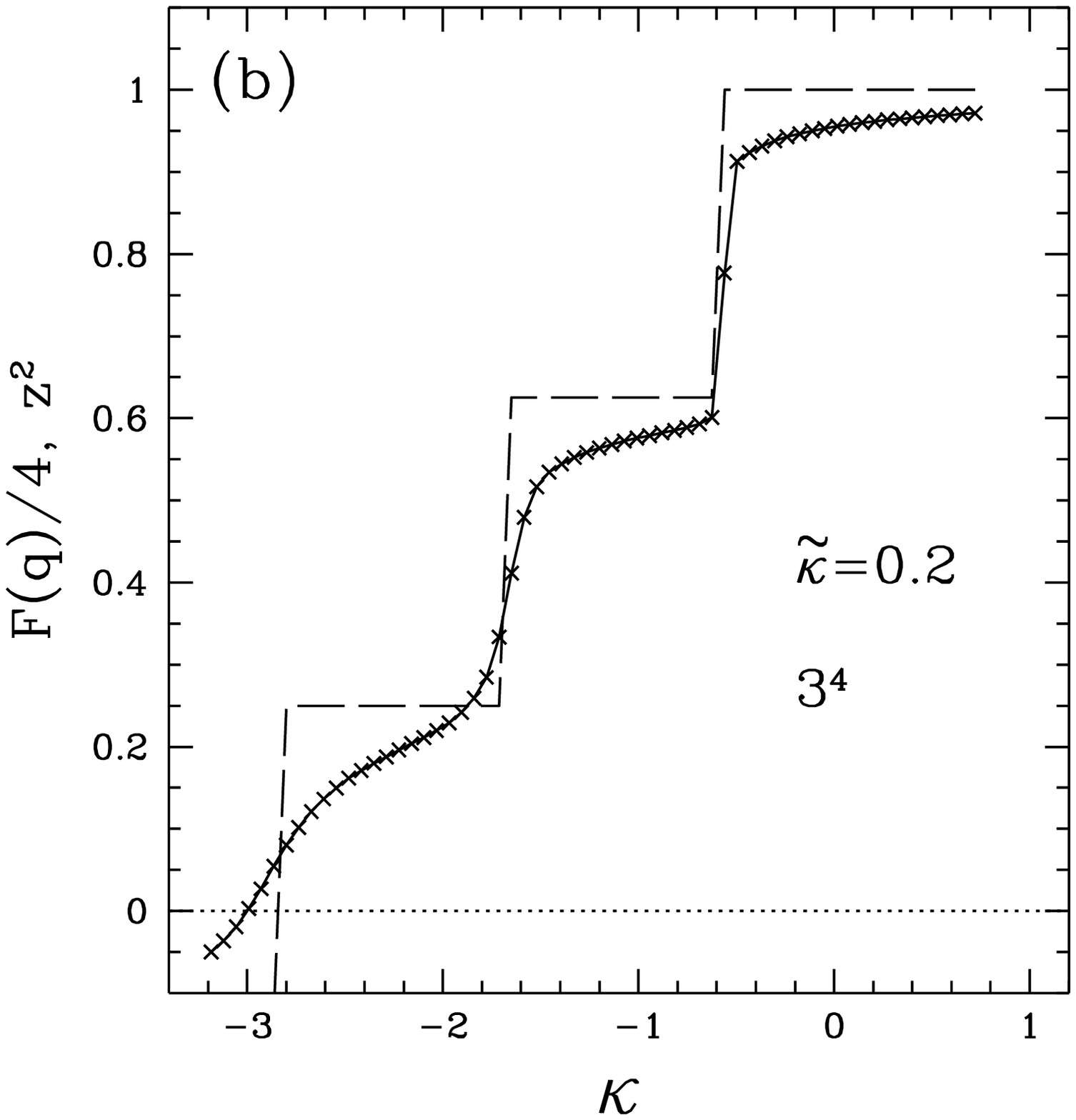}     \\
\end{tabular}
\vspace*{-0.5cm}
\caption{ \noindent {\em  
The quantity $\quart F(q)$ (dashed line),  obtained from 
the minimization of the classical action ({\protect \eq{SCLASSICAL}}),
and the internal energy density $z^2$ (crosses), obtained from the 
numerical simulation, as a function of $\k$ for $\tk=0.2$
on $4^4$ (fig.~a) and $3^4$ (fig.~b) lattices. 
}}
\label{FIGFQ}
\end{figure}

In this subsection we compare the simulation results 
for $v$ in the FM and $v_{{\rm H}}$ in the FMD phase  
with the perturbative formulas which we derived in Sect.~\ref{PERT}. 
In fig.~\ref{TFMD} we have plotted once more the $v$ and $v_{{\rm H}}$ data 
of fig.~\ref{SCANS}a which were obtained at $\tk=0.2$ on a 
$4^4$ lattice. The circles in fig.~\ref{TFMD} were obtained by evaluating  
the one-loop formula for $v_{{\rm H}}$ (\eq{PMAG2}) 
numerically on both sides of the FM-FMD phase transition  on the same 
lattice and at the same values of $\k$ where we performed the      
numerical simulations. The phases $q(\tk,\k)$ at a given value 
of $\k$ were determined analytically by minimizing the classical action 
in eq.~(\eq{SCLASSICAL}) with $L=4$. 
Fig.~\ref{TFMD} shows that the numerical results 
are nicely reproduced by the analytic formula (\eq{PMAG2}) at large values 
of $\k$. At smaller $\k$ values the deviations start to become larger. 
While we do not understand this phenomenon in detail, we believe that  
it may be related to the metastabilities mentioned in Sect.~\ref{NUM_PD}. 
(Higher orders in perturbation theory could also be sizable though.)
The jumps at which the components of $q$ condense occur slightly delayed.
This distorts the $\k$-dependence of the propagator, which depends 
also explicitly on $\k$ and not only implicitly 
through the phases $q(\tk,\k)$. 
{}From the minimizing phases $q(\tk,\k)$ on the $4^4$ lattice
we have computed the function $F(q)$, cf. eq.~(\eq{FQ}), which we plotted  
in fig.~\ref{FIGFQ}a (dashed line) versus $\k$. The numerical results for 
$z^2$, which in the mean-field approximation are proportional to $F(q)$ 
are represented in this graph by the crosses. 
It can be seen that the gap between the two curves 
widens up when $\k$ is lowered, which is presumably due to the systematic 
delay of the transition events in the numerical simulation. We should observe 
a smaller shift on a smaller lattice if this scenario is correct. Fig.~\ref{FIGFQ}b
shows that the gap between the two curves shrinks indeed on the $3^4$
lattice. We also checked that the numerical metastabilities on a $5^4$ lattice 
get stronger making the agreement with the analytic results worse. Note that 
the discrepancies between perturbation theory and numerical data 
occur at the same locations in figs.~\ref{TFMD} and \ref{FIGFQ}a.

In the following we will discuss only the FM phase. 
The numerical metastabilities mentioned in the 
previous paragraph have an effect on the simulation results 
only in the near vicinity of the FM-FMD phase transition. 
\begin{figure}
\centerline{
\epsfxsize=14.0cm
\vspace*{0.5cm}
\epsfbox{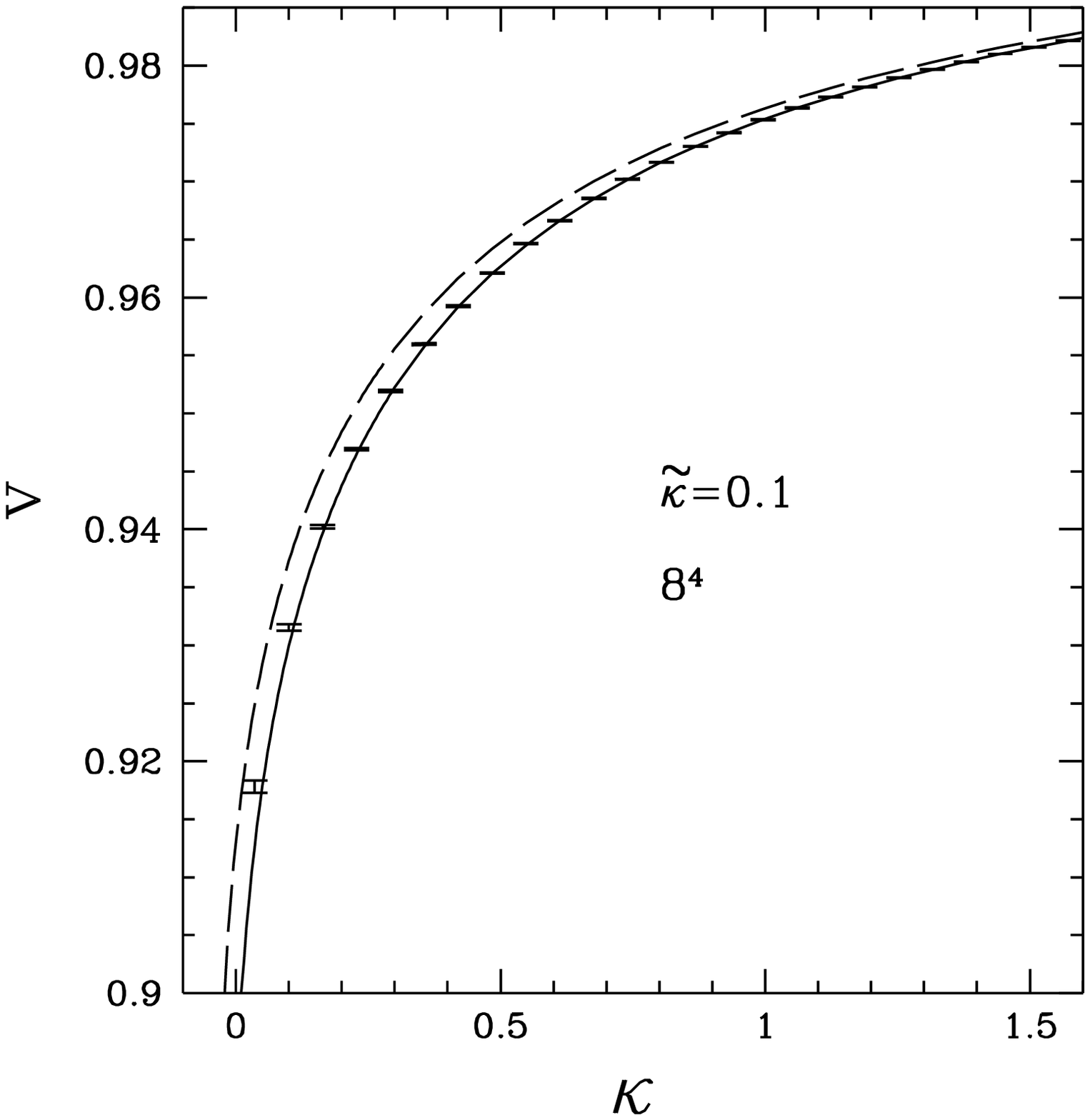}
}
\vspace*{-1.2cm}
\caption{ \noindent {\em  
The magnetization $v$ as a function of $\k$ for $\tk=0.1$. The lattice size 
is $8^4$. The one-loop and two-loop results for $v$ are represented by the dashed, 
cf.~eq.~({\protect \eq{PMAG2}}), and solid, cf.~eq.~({\protect \eq{PMAG4}}),
curves. 
}}
\label{MAG01}
\end{figure}

In fig.~\ref{MAG01} the magnetization $v$ for $\tk=0.1$ is displayed 
as a function of $\k$. The lattice size is $8^4$. 
At each $\k$ point we have accumulated a statistics of         
$10^5$ Metropolis sweeps. The magnetization was measured after each sweep 
and its error was computed by means of eq.~(\eq{ER}). 
To compare the numerical data with the perturbative formulas, we have 
numerically evaluated the integrals (replaced by lattice sums)  
in eq.~(\eq{PMAG2}) and (\eq{PMAG4}) 
for a large number of $\k$ values in the FM phase on the same lattice which 
we used in the numerical simulations.  The obtained results for $v$
in eqs.~(\eq{PMAG2}) and 
(\eq{PMAG4}) are represented by the dashed and solid curves. 
Fig.~\ref{MAG01} shows that the two-loop 
formula (\eq{PMAG4}) provides, as expected,  a much better 
description of the numerical 
data than the one-loop formula (\eq{PMAG2}).  
The fact that perturbation theory in $1/\tk$ 
remains valid down to such small 
values of $\tk$ is because the actual expansion parameter is not $1/\tk$ 
but $1/(16 \p^2 \tk)$ where the factor $1/(16 \p^2 )$ comes from    
the loop integrals.
\begin{figure}
\centerline{
\epsfxsize=14.0cm
\vspace*{0.5cm}
\epsfbox{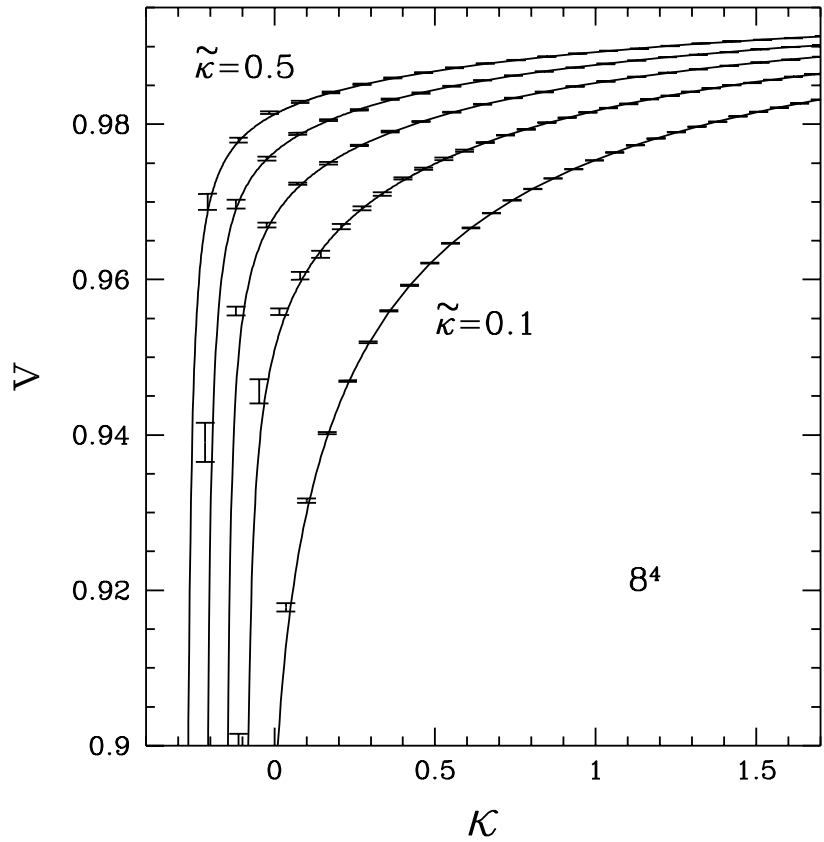}
}
\vspace*{-1.2cm}
\caption{ \noindent {\em  
The magnetization $v$ as a function of $\k$ for 
several values of $\tk$ . The five curves correspond, from the bottom to the top, 
to $\tk=0.1$, $0.2$, $0.3$, $0.4$ and $0.5$. The lattice size 
is in all cases $8^4$. The solid lines were obtained by evaluating expression 
({\protect \eq{PMAG4}}) for a large number 
of $\k$ values numerically.                                      
}}
\label{MAG_KAPPA}
\end{figure}

The $\tk$-dependence of the  magnetization $v$ is shown in fig.~\ref{MAG_KAPPA}, 
where we have plotted the magnetization data for five different $\tk$ values 
versus $\k$. The lattice size is also in this case 
$8^4$ and the statistics at each $\k$ point the same as in fig.~\ref{MAG01}.
In order to better monitor the drop of the magnetization  near 
the FM-FMD phase transition we have increased the density of points in that 
region. The solid lines represent again the perturbative result 
according to eq.~(\eq{PMAG4}). The agreement between the numerical 
data and the analytic curve is in all cases excellent.
The graph shows that the drop to the critical point is becoming 
steeper when $\tk$ is increased. This phenomenon is a consequence of the fact 
that the critical exponent $\eta$ in eq.~(\eq{EETA}) 
decreases with increasing $\tk$.  

\begin{figure}
\centerline{
\epsfxsize=14.0cm
\vspace*{0.5cm}
\epsfbox{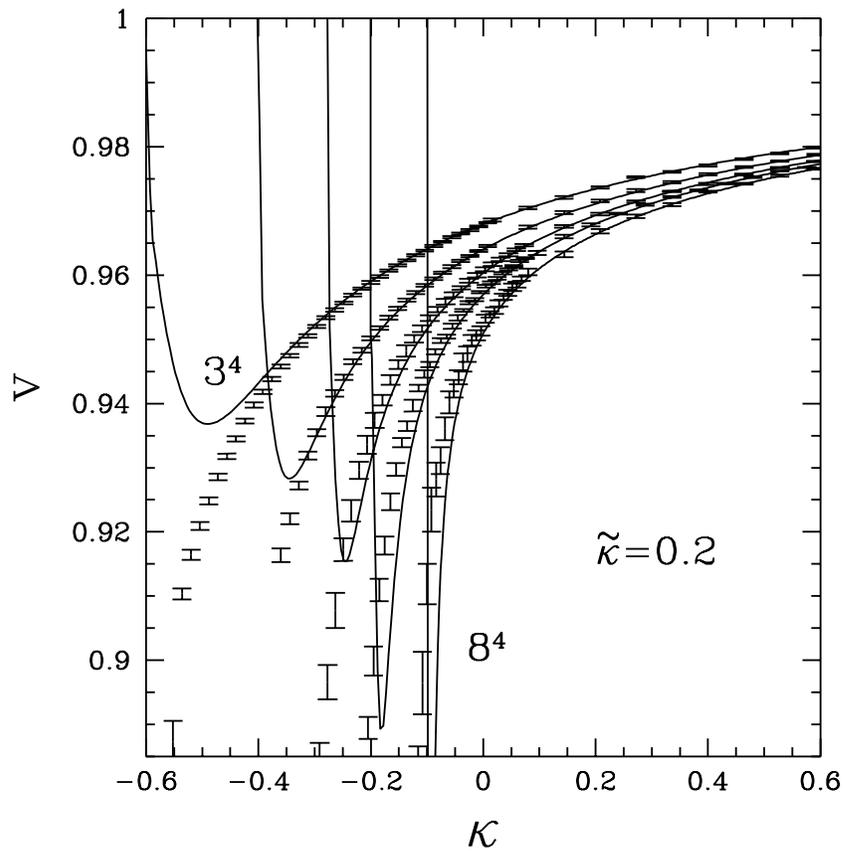}
}
\vspace*{-1.2cm}
\caption{ \noindent {\em  
The magnetization $v$ versus $\k$ for $\tk=0.2$ and several values of $L$. 
The five curves correspond, from the left to the right, 
to $L=3$, $4$, $5$, $6$ and $8$. 
The five solid lines were obtained by computing            
({\protect \eq{PMAG4}}) numerically on lattices of the same size.        
}}
\label{MAG_L}
\end{figure}

Finally we demonstrate that also 
the volume dependence of the magnetization data in the FM phase, 
cf. fig.~\ref{v_fs}, is nicely reproduced 
by the perturbative formula (\eq{PMAG4}). In fig.~\ref{MAG_L}
we have plotted $v$ as a function of $\k$ for $\tk =0.2$ and five different lattices 
ranging in size from $3^4$ to $8^4$. The statistics at each $\k$ value 
is about $10^5 \times (8/L)^2$ Metropolis sweeps. The solid lines represent 
again the perturbative results according to eq.~(\eq{PMAG4}). The two-loop  curve 
agrees nicely with the numerical data down to the value of $\k$ 
where the analytic    
curve has a  minimum, but starts to deviate when $\k$ is lowered beyond that 
value. The two-loop curve increases while the numerical data continue to fall off.
This shows that the two-loop formula (\eq{PMAG4}) 
is valid only in the $\k$ interval above 
the minimum. The self-energy $\S(k)$ in eq.~(\eq{PMAG5}) diverges at 
\be
\k = \k_{{\rm min}}=- 2\;\tk \;  (1 - \cos 2\pi/L ) \; ,
  \lb{RANGE}
\ee
because the inverse tree-level propagator $\D_0^{-1}(k)$ has a zero eigenvalue 
(for some nonzero $k$)
at this value of $\k$. This implies that $v$ in eq.~(\eq{PMAG4}) approaches 
one in the limit $\k \searrow \k_{{\rm min}}$.
(Lowering $\k$ beyond this value would lead to a negative eigenvalue of
the tree-level inverse propagator, and this instability causes $q$
to condense to the smallest possible value, cf. eq.~(\eq{QDIS}).)
The plot shows that the minimum of the two-loop curve gets smaller and also 
narrower when the size of the lattice is increased. The minimum drops 
to zero in the infinite volume limit at the $\k$-value where $a(\tk, \k)$
vanishes, cf. eq.~(\eq{XKAPPA_C}). 
This $\k$-value coincides with the one-loop estimate (\eq{KAPPA_C}) only 
in the limit $\k \ra \infty$. At $\tk=0.2$ we find that $v$ 
drops to zero at $\k \approx 0.03446$ which is by about 15\% larger than the 
one-loop estimate in eq.~(\eq{KAPPA_C}).   

{}From eq.~(\eq{RANGE}) we find that for $\tk=0.2$, $\k_{{\rm min}}=-0.6$, $-0.4$, 
$\approx -0.2764$, $-0.2$ and $-0.1$ for $L=3$, $4$, $5$, $6$ and $8$. 
Fig.~\ref{MAG_L} shows that the two-loop curve approaches one at these
values of $\k$. 

\section{Summary and Outlook}
\lb{OUT}
In this paper we have calculated the phase diagram of the {\em reduced model}
for a gauge-fixed U(1) lattice gauge theory 
in the mean-field approximation. The phase diagram contains a 
ferromagnetic (FM), anti-ferromagnetic (FM), paramagnetic (PM) and, as a novelty, 
also a ferromagnetic directional (FMD) phase, where not only the U(1) symmetry 
is broken but also the vector field condenses.
The mean-field results for the phase diagram were confirmed by numerical 
simulations. 
The continuum limit of the model corresponds to a continuous phase
transition between  the FM phase and the FMD phase. 
We have studied the nature of this phase transition 
in lattice perturbation theory and demonstrated 
that the magnetization vanishes in the continuum limit, because of 
infra-red effects, and that the global U(1) symmetry gets 
restored in this limit. This phenomenon cannot be 
understood in the mean-field approximation.
We have shown that the numerical data for the 
magnetization are in good agreement with the results of the weak 
coupling expansion in the FM phase. 
We have calculated the critical coupling of the FM-FMD 
phase transition at large values of $\tk$ 
both in the mean-field approximation and in perturbation theory, and 
find in both cases a small positive value. At $\tk \ra \infty$ 
the mean-field result for the FM-FMD phase transition 
is smaller by about a factor two than the perturbative result.

As a next step we will take up again various 
proposals for lattice chiral gauge theories and investigate whether the  
problems associated with the strongly fluctuating  gauge degrees 
of freedom can be overcome by gauge fixing.
In refs. \cc{BoGoSh97b} and \cc{BoGoSh97c} we will show for the case of the 
reduced abelian Wilson-Yukawa (Smit-Swift) model that 
a) the species doublers decouple in the continuum limit,   
and b) that the fermion spectrum contains only the desired states, 
namely a massless charged left-handed fermion that couples to the gauge field 
and a massless neutral right-handed fermion that decouples.          
We expect to find similar positive results 
also for other fermion formulations, using a
Majorana-Wilson term instead of a Dirac-Wilson term \cc{Pr91}, 
domain wall fermions with 
waveguide \cc{GoJa94}, or staggered fermions \cc{Sm92}.                       

It is challenging to study the U(1) model with gauge fields turned on.
It should be possible to determine the fermion spectrum in 
the Coulomb phase and see if it remains unaffected at small values 
of the gauge coupling. A change of the fermion spectrum 
should manifest itself as a new phase transition in the fermion sector.

It is also important to extend the 
gauge-fixing approach to nonabelian gauge theories. 
This implies that we first have 
to specify how to discretize and simulate
the ghost part of the action (\eq{CACTION}). 
The nonabelian case is very interesting  
because in this case we can ask whether 
confinement emerges at small values of the gauge coupling. 
\subsubsection*{Acknowledgements}
We thank M. Ogilvie for discussions. 
W.B. is supported by the Deutsche Forschungsgemeinschaft (DFG). M.G.   
is supported by the US Department of Energy as an Outstanding Junior Investigator,
and Y.S. is supported in part by the US-Israel Binational Science Foundation, 
and the Israel Academy of Science. The numerical calculations 
were performed on the SP2 at DESY-IfH Zeuthen and numerous workstations 
and PCs at the Physics Departments of Washington 
University, St. Louis, and Humboldt University, Berlin. Some of the
first explorative calculations were performed on the HP-cluster of the Center
for Computational Physics of the University of Tsukuba. W.B. thanks  
Washington University, St. Louis for hospitality and M.G. thanks 
the Center for Computational Physics of the University of Tsukuba,
the Benasque Center for Physics and Humboldt University, Berlin, for 
hospitality. 

\end{document}